\documentclass[fleqn,usenatbib]{mnras}

\usepackage{xparse}
\usepackage{bm,upgreek}
\usepackage{tabularx}
\usepackage{amssymb}

\usepackage[T1]{fontenc}

\DeclareRobustCommand{\VAN}[3]{#2}
\let\VANthebibliography\thebibliography
\def\thebibliography{\DeclareRobustCommand{\VAN}[3]{##3}\VANthebibliography}

\usepackage{graphicx}	
\usepackage{amsmath}	
\usepackage{amssymb}	
\usepackage[capitalise]{cleveref}
\usepackage[dvipsnames]{xcolor}

\newcommand{\vth}{\boldsymbol{\theta}}

\newcommand{\vect}[1]{\boldsymbol{#1}}

\newcommand{\msv}{\citetalias{Monsalve2017}}
\newcommand{\np}{\citetalias{Bowman2018}}

\newcommand{\model}{M}
\newcommand{\fixedmodel}{\textit{F}}
\newcommand{\param}{P}

\newcommand{\choice}{\textit{C}}

\newcommand{\meas}{\textbf{X}}
\newcommand{\both}{\textbf{T}}

\ExplSyntaxOn
\NewDocumentCommand\matr{m}
 {
  \commexo_vector:n { #1 }
 }

\cs_new_protected:Npn \commexo_vector:n #1
 {
  \tl_map_inline:nn { #1 }
   {
    \commexo_vector_inner:n { ##1 }
   }
 }

\cs_new_protected:Npn \commexo_vector_inner:n #1
 {
  \tl_if_in:VnTF \g_commexo_latin_tl { #1 }
   {
    \mathbf { #1 } 
   }
   {
    \tl_if_in:VnTF \g_commexo_ucgreek_tl { #1 }
     {
      \bm { #1 } 
     }
     {
      \tl_if_in:VnTF \g_commexo_lcgreek_tl { #1 }
       {
        \commexo_makeboldupright:n { #1 }
       }
       {
        #1 
       }
     }
   }
 }

\cs_new_protected:Npn \commexo_makeboldupright:n #1
 {
  \bm { \use:c { up \cs_to_str:N #1 } }
 }

\tl_new:N \g_commexo_latin_tl
\tl_new:N \g_commexo_ucgreek_tl
\tl_new:N \g_commexo_lcgreek_tl
\tl_gset:Nn \g_commexo_latin_tl
 {
  ABCDEFGHIJKLMNOPQRSTUVWXYZ
  abcdefghijklmnopqrstuvwxyz
 }
\tl_gset:Nn \g_commexo_ucgreek_tl
 {
  \Gamma\Delta\Theta\Lambda\Pi\Sigma\Upsilon\Phi\Chi\Psi\Omega
 }
\tl_gset:Nn \g_commexo_lcgreek_tl
 {
  \alpha\beta\gamma\delta\epsilon\zeta\eta\theta\iota\kappa
  \lambda\mu\nu\xi\pi\rho\sigma\tau\upsilon\phi\chi\psi\omega
  \varepsilon\vartheta\varpi\varphi\varsigma\varrho
 }

\ExplSyntaxOff

\defcitealias{Monsalve2017}{M17}
\defcitealias{Bowman2018}{B18}

\usepackage{newtxtext,newtxmath}

\title[EDGES Bayesian Calibration]{A Bayesian Calibration Framework for EDGES}

\author[S.~G.~Murray et al.]{
Steven G. Murray,$^{1}$\thanks{E-mail: steven.g.murray@asu.edu}
Judd D. Bowman,$^{1}$
Peter H. Sims,${^2}$
Nivedita Mahesh,$^{1}$
Alan E. E. Rogers,$^{3}$ \newauthor
Raul A. Monsalve,$^{4,1,5}$
Titu Samson,$^{1}$
Akshatha Konakondula Vydula,$^{1}$
\\
$^{1}$School of Earth and Space Exploration, Arizona State University, Tempe, AZ 85287, USA\\
$^{2}$Department of Physics and McGill Space Institute, McGill University, Montr\'eal, QC H3A 2T8, Canada\\
$^{3}$Haystack Observatory, Massachusetts Institute of Technology, MA 01886, USA\\
$^{4}$Space Sciences Laboratory, University of California Berkeley, Berkeley, CA 94720, USA\\
$^{5}$Facultad de Ingenier\'ia, Universidad Cat\'olica de la Sant\'isima Concepci\'on, Alonso de Ribera 2850, Concepci\'on, Chile\\
}

\date{Accepted XXX. Received YYY; in original form ZZZ}

\pubyear{2020}

\begin{document}

\renewcommand{\equationautorefname}{Eq.}
\renewcommand{\sectionautorefname}{Sec.}
\renewcommand{\subsectionautorefname}{Sec.}
\renewcommand{\figureautorefname}{Fig.}

\label{firstpage}
\pagerange{\pageref{firstpage}--\pageref{lastpage}}
\maketitle

\begin{abstract}
We develop a Bayesian model that jointly constrains receiver calibration, foregrounds and cosmic 21\,cm signal for the EDGES global 21\,cm experiment.
This model simultaneously describes calibration data taken in the lab along with sky-data taken with the EDGES low-band antenna.
We apply our model to the same data (both sky and calibration) used to report evidence for the first star formation in 2018.
We find that receiver calibration does not contribute a significant uncertainty to the inferred cosmic signal ($< 1\%$), though our joint model is able to more robustly estimate the cosmic signal for foreground models that are otherwise too inflexible to describe the sky data. 
We identify the presence of a significant systematic in the calibration data, which is largely avoided in our analysis, but must be examined more closely in future work.
Our likelihood provides a foundation for future analyses in which other instrumental systematics, such as beam corrections and reflection parameters, may be added in a modular manner.
\end{abstract}




\section{Introduction}
\label{sec:intro}

The globally-averaged brightness temperature of the hyperfine spin-flip transition of neutral hydrogen (the 21\,cm line) is a powerful probe of the thermal history of the early Universe \citep[$z\sim 6-30$; for reviews, see eg.][]{Furlanetto2006,Pritchard2012,Furlanetto2016}.
Accurately observing this brightness temperature, and separating it from the bright foreground emission of our Galaxy, have proven to be an exceptional challenge. 
Several instruments have taken up this challenge, including EDGES \citep{Bowman2008,Rogers2012}, LEDA \citep{Bernardi2016}, BIGHORNS \citep{Sokolowski2015}, and SARAS \citep{Girish2020}.
Since the publication of the first evidence for star formation in Cosmic Dawn by the EDGES collaboration \citep[][hereafter \np]{Bowman2018}, there has been an increased interest in independent verification, resulting in several new and upcoming experiments, eg. SARAS3 \citep{Nambissan2021}, ASSASSIN \citep{Mckinley2020} and REACH \citep[eg.][]{Anstey2020}.
Importantly, the recent results of SARAS3 \citep{Singh2022} appear inconsistent with the inferred cosmic signal of \np, suggesting that either measurement (or both) may be contaminated by systematics.

Despite the overwhelming magnitude of the foregrounds \citep[$\sim10^5$ times the signal, eg.][]{Shaver1999}, they are not the primary challenge \textit{in isolation}. 
Indeed, physical models of foreground spectra are incredibly smooth, defined by relatively low-order deviations from a power-law \citep{Jelic2010}. 
Conversely, most cosmological signals have more rapid spectral structure, much of which is not captured by these same low-order basis-sets \citep{Bevins2021}.
Rather, the primary challenge arises via the multiplication of these bright smooth foregrounds by relatively small spectral structures induced by the instrument.
These come from a number of physically distinct mechanisms, including \textit{beam chromaticity} (in which angular structure in the sky and beam are translated to frequency structure via the frequency-dependence of the beam-shape primarily due to reflections from the edges of the ground plane and nearby objects (Rogers et al., 2022, \textit{submitted}), reflection parameters of the signal chain, and receiver gains.
Structures created by these instrumental systematics that happen to share similar scales as the expected signal must thus be avoided, or calibrated to a precision of about $10^{-5}$ in order to provide minimal contamination to the estimated cosmic signal.

The results of \np\ were carefully calibrated, and are expected to have residual systematics that are subdominant to the (surprisingly strong) cosmological absorption feature.
Nevertheless, the analysis was performed in a way that obscures the relationship between the uncertainties on the known systematics and the final uncertainties on the cosmological estimate. 
That is, the reported error bars were obtained via independent estimates of the propagated uncertainties of various systematics, added in quadrature. 
This is a crude estimate, not accounting for correlations in the effects of different unknown parameters, nor properly accounting for our prior knowledge of these parameters.

This is made compelling by \citet{Sims2020}, who use the Bayes Factor -- a rigorous metric of the comparative evidence for one model over another -- to argue that a simple unmodeled systematic that exhibits as a damped sinusoid in the spectrum would disfavor a strong absorption feature \citep[see eg.][for prior studies that suggested a similar phenomenological systematic]{Hills2018,Singh2019}.
Nevertheless, this statement is highly dependent on our prior knowledge; we know of no physical systematic that should arise as such a simple damped sinusoid. 
On a more nuanced view, it is possible that the combination of receiver gains, reflections and beam chromaticity would compound to yield a systematic with  approximately sinusoidal structure; however, it then becomes important to know what the expected amplitude and period of such a sinusoid might be, and how likely it is (given the physical uncertainties of the parameters involved) that it would reach the strength and shape required to obviate the cosmological feature.

To properly address these questions, we require a full Bayesian forward-model. Such a model begins with the unknown physical parameters, for which we have reasonable estimates of uncertainty, and propagates those uncertainties self-consistently all the way through to the final signal estimation. 
This captures the full correlated, non-Gaussian probability distributions of the unknown parameters, allowing a more rigorous determination of their marginalised uncertainty. 
It also allows for comparing different models.

There is precedent for Bayesian models in global 21\,cm experiments. Besides the use of Bayesian techniques to determine posteriors on the signal and foreground parameters \citep[eg.][]{Monsalve2017a,Monsalve2018,Monsalve2019,Singh2019,Sims2020,Bevins2022}, there has been work on including various systematics in forward models, predominantly led by the REACH collaboration.
This pioneering work encompasses foreground models and beam chromaticity \citep{Anstey2020}, antenna models \citep{Anstey2022}, generalized systematics \citep{Scheutwinkel2022a} and non-Gaussian noise statistics \citep{Scheutwinkel2022}. 
Perhaps most relevant for this work, \citet{Roque2020} considers receiver calibration under a Bayesian model-selection framework.
In this work, the focus was on modeling the receiver gain posteriors, in order to determine a posterior on the calibrated temperature (and also to choose the number of polynomial terms required for calibration in a self-consistent way). 
To do this efficiently, \citet{Roque2020} use the method of conjugate priors, yielding an analytic solution to the posterior.
In this paper, we implement a very similar Bayesian model for the receiver gains; however, we do not adopt the conjugate prior formalism, despite its efficiency. 
We do this because beyond the receiver gains, we are interested in simple models for the reflection coefficients. The required flexibility for these extended models makes using conjugate priors more difficult. 
Furthermore, we extend the forward model through to \textit{joint analysis} of the receiver gain, foregrounds and cosmic signal.

This paper is the beginning of a larger project in which the entirety of the EDGES analysis chain is to be cast in a Bayesian forward-model. 
In this paper, we focus purely on receiver calibration. 
As this paper is primarily about the \textit{technique}, we will apply the model to the data that constituted the result of \np.
As such, this paper is not intended to form a full `validation' of the \np\ result; rather, it provides a necessary step in building confidence that certain systematics (in receiver gains) are unlikely to have caused the surprising results previously obtained. 
A more complete verification requires the modeling of all \textit{known} systematics, and furthermore an expansion of the data (preferably to independent telescopes) to investigate potential \textit{unknown} systematics.

The layout of the paper is as follows. \S\ref{sec:data} describes the data used throughout the paper. \S\ref{sec:math-mcmc} introduces Bayesian inference in general, and derives a high-level likelihood for global experiments. \autoref{sec:calibration-model} dives into the details of the probabilistic receiver gain calibration model adopted in this paper. \S\ref{sec:field-data-model} extends this probabilistic model to include sky data, before we analyse that data with our Bayesian model in \autoref{sec:results}. Finally, we summarize and conclude in \autoref{sec:conclusions}.

All analysis in this work is open-source and available in Jupyter notebooks and Python scripts\footnote{Available at \url{https://github.com/edges-collab/bayesian-calibration-paper-code/releases/tag/submitted}}.

\section{Data Used}
\label{sec:data}
All data used in this paper comes from \np.
Only part of the data required here was made publicly available by \np, namely the time-averaged sky spectra. 
In addition to the sky spectrum itself, we require several calibration products in this paper, for which we use the exact data/settings used in \np.

The sky spectrum consists of observations between day 250 of 2016 through to day 98 of 2017 (138 days after initial data quality cuts). 
Each integration from each night is filtered for RFI and other systematic outliers, including cuts on metadata such as local humidity and potential saturation of the analog-to-digital-converter (ADC).
After filtering a day's worth of data, all integrations within the 12 hours of LST corresponding to the galactic centre being below the horizon are averaged together.
This averaged spectrum is then calibrated for the receiver gain, beam correction and path losses.
Further filtering is performed on the calibrated, averaged spectrum from each night, checking for outliers.
Finally, the 138 days are averaged together, and the spectra are binned in frequency bins of $\sim 0.390\,{\rm MHz}$.
This final spectrum is referred to as $\hat{\overline{T}}_{\rm sky,bc}$ in this paper (cf. Eq. \ref{eq:processed_q}), and we directly use the publicly available data\footnote{Available at \url{https://www.nature.com/articles/nature25792/figures/1}} in this paper.
We refer the interested reader to \np\ for details on the data analysis.

In this paper, we often need to `undo' the calibration of the fiducial dataset described above, in order to recalibrate with different parameters. 
This process is defined in Eq. \ref{eq:processed_q}, and requires the nominal receiver calibration ($\hat{T}_0^{\rm ant}$ and $\hat{T}_1^{\rm ant}$), beam correction and path loss.

The path loss is, in general, a product of antenna, balun, connector and ground losses. The antenna loss is produced via simulation with FEKO \citep{Elsherbeni2014}. In the case of the public data from \np, the ground loss is set to unity (i.e. ignored).
The beam correction is produced via Eq. \ref{eq:beamcorr} \citep[cf.][]{Mozdzen2019}, using the Haslam all-sky map \citep{Haslam1982} with a spatially-invariant spectral index of -2.5, along with a beam model produced with FEKO \citep{Mahesh2021}. 
While these basic products are not publicly available, we show their final form in Fig. \ref{fig:losses}.

To obtain the receiver calibration we directly use outputs of the original C-code adopted in \np\footnote{This code, with scripts to run it with the same settings as \np, is available at \url{https://github.com/edges-collab/alans-pipeline}}. 
This includes frequency-dependent values of five receiver-calibration coefficients as well as the reflection coefficients of the receiver and antenna. 
These calibration parameters and reflection coefficients allow us to \textit{de}-calibrate the public data (essentially taking it back to its raw form\footnote{As described further in the paper, we do this de-calibration, instead of starting directly from the raw data, because we wish to retain the exact averaging and flagging used on the public data, without re-performing this compute-heavy task.}). 
However, to \textit{re}-calibrate the data requires the calibration measurements used to initially derive these calibration solutions, along with the various original settings used in the analysis.
These calibration measurements were taken in the lab in September 2015. The observation includes the simultaneously measured spectra and temperatures from the four input `calibration' sources (ambient, hot load, open and shorted long cable) plus an `antenna simulator' designed to mimic the reflection coefficients of the antenna, as well as reflection coefficients for these input sources and the internal switch and receiver, and measurements of the resistance of the (SOL) calibration standards used to measure the reflection coefficients. 
These calibration measurements are here analysed with a new publicly-available calibration code, \textsc{edges-cal}\footnote{Available at \url{https://github.com/edges-collab/edges-cal}} to produce the five calibration parameters referenced previously.
A detailed report of the use of the new code to produce the calibration parameters in this paper is available in \citet{Murray2022a}, which also demonstrates slight variations in the results between the codes -- even with (nominally) the same input settings.
The largest difference concerns the modelling of the various reflection parameters required (one for each calibration source, plus the internal switch, receiver and antenna). 
Since we are not concerned in this paper with re-modelling the reflection parameters, we simply take the direct output of the code used in \np\ and input those values to our own calibration using \textsc{edges-cal}\footnote{In this work, we use the full suite of new EDGES pipeline codes, all open-source and available at \url{https://github.com/edges-collab}. Specifically, we use \textsc{read-acq} v0.5.0, \textsc{edges-io} v4.1.3, \textsc{edges-cal} v6.2.3, \textsc{edges-analysis} v4.1.3 and \textsc{edges-estimate} v1.3.0}.

\section{Mathematical and Bayesian Framework}
\label{sec:math-mcmc}

\subsection{Notational Preliminaries}
\label{sec:model:notation}

Throughout, bold upright quantities, eg. $\matr{Q}$, will refer to matrices, and bold italic, eg. $\vect{q}$ will refer to vectors (where possible, vectors will also be lower case). 
Throughout, the symbol `$\circ$' will refer to Hadamard (i.e. element-wise) multiplication, and $\oslash$ will refer to Hadamard division. The symbol $\matr{I}_n$ will refer to the $n\times n$ identity matrix.
We will construct (row) vectors using square brackets surrounding elements separated by commas, eg. $\vect{q}_{\rm cal} = \left[ q_1, q_2, \dots \right]$, and assume that the transpose of a row vector, $\vect{q}^T$, is a column vector.

The ensemble average of a random variable will be denoted by angle brackets, eg. $\langle q \rangle$, while a sample mean will be denoted by an over-bar, eg. $\bar{q}$.
An estimate of a quantity will be denoted by a hat, eg. $\widehat{q} = \bar{q}$.

When denoting parameters that are statistics of a certain observable which itself is measured for multiple independent sources (eg. the variance, $\sigma^2$ of the three-position-switch ratio, $q$ for the open cable), we will denote the the observable as eg. $q_{\rm open}$, but will `lift' the source by one subscript level when denoting the statistics, i.e. $\sigma^2_{\rm q, ant}$ rather than $\sigma^2_{q_{\rm ant}}$.

Throughout, we will use curly braces surrounding elements separated by commas to construct \textit{sets}. Symbols denoting such sets will typically be in upper-case Latin calligraphic font, eg. $\mathcal{T}_{\rm NW} = \{{\rm cos, unc, sin}\}$, and usually these will denote sets of \textit{labels} (eg. the three labels associated with `noise-wave' temperatures). Sets of quantities associated with these labels may be represented in the short-hand notation $T_{\mathcal{T}_{\rm cal}} \equiv \{T_p\ |\ p \in \mathcal{T}_{\rm cal}\}$.


\subsection{Bayes' Theorem}
Bayesian approaches to parameter inference and model selection have become extremely popular in the astrophysics and cosmology literature. 
As such, we will only describe them briefly, referring the interested reader to more in-depth resources, such as \citet{Jaynes2003}.

Bayesian statistics is fundamentally the update of the \textit{credence} in a certain model given the acquisition of new data pertaining to the model. 
That is, it presupposes an existing credence (the ``prior'') and some observations, and given a likelihood of obtaining those observations given the parameters of the model, it yields an updated credence. This process is described by Bayes' formula:
\begin{equation}
    P(\vth|D, \mathcal{M}) = \frac{\mathcal{L}(D|\vth, \mathcal{M})\pi(\vth)}{\mathcal{Z}(D|\mathcal{M})}.
\end{equation}
The `model', $\mathcal{M}$, is here parameterized by the set of parameters $\vth$\footnote{In principle, the total possible set of models may contain completely different parameterizations. In practice, we typically explore a single parameterization, $\mathcal{M}$, at a time, which has parameters $\vth$.}. The LHS represents the `posterior' credence of $\vth$ under model $\mathcal{M}$ \textit{after} observing data $D$, while the RHS takes the `prior' credence, $\pi(\vth)$, and updates it with the `likelihood' of the data, $\mathcal{L}(D|\vth)$, normalized by the `evidence', $\mathcal{Z}(D|\mathcal{M})$.

Typically (although see \citet{Roque2020}) the evidence is impossible to write down analytically, but may be computed as the integral of the likelihood over the prior subspace. 
In this paper, we use the \textsc{polychord} sampler which is able to provide not only samples from the posterior, but an estimate of the evidence, $\mathcal{Z}$.

\subsection{The Gaussian Likelihood}
\label{sec:gaussian_likelihood}
In this paper, we will exclusively use a \textit{Gaussian} likelihood. In this likelihood, we model the data as being sampled from a multivariate Gaussian distribution with mean vector $\vect{\mu}(\vth) \equiv \mu(\vth, \vect{x})$ and covariance matrix $\matr{\Sigma}(\vth) \equiv \Sigma(\vth, \vect{x})$.
The mean vector is typically dependent both on the parameters of the model and a predicate variable, $\vect{x}$, which for this paper will be taken to be known with certainty (typically it will be frequency and/or input source). The data is taken to be sampled at particular values of this predicate variable, $\vect{d} \sim \mathcal{N}\left(\vect{\mu}(\vth), \matr{\Sigma}(\vth)\right)$.

The likelihood is thus given by
\begin{equation}
    \label{eq:gaussian-likelihood}
    \mathcal{L}_g(\vect{d} | \vth) \propto \sqrt{|\matr{\Sigma}^{-1}|} \exp\left\{- \vect{r}^T \matr{\Sigma}^{-1} \vect{r} \right\},
\end{equation}
where the model residual is given by
\begin{equation}
    \vect{r} = \vect{d} - \vect{\mu}(\vth).
\end{equation}

Note that the Gaussian likelihood is valid so long as the residuals, $\vect{r}$ are Gaussian distributed. Given that some raw data $\vect{d}$ is Gaussian distributed, the linearly transformed data $\vect{d}' = \matr{A}\vect{d} + \vect{b}$ is also Gaussian distributed. Here the $\vect{b}$ is simply absorbed into $\vect{\mu}$, but the scaling matrix $\matr{A}$ results in a scaled covariance. 
Thus, in general with raw Gaussian-distributed data $\vect{d}$ with covariance $\matr{\Sigma}_{\vth}$ we may write the likelihood as 
\begin{equation}
    \label{eq:general-gaussian-likelihood}
    \mathcal{L}_g(\vect{d} | \vth) \propto \sqrt{|\matr{\Sigma}_A^{-1}|} \exp\left\{- \vect{r'}^T \matr{\Sigma}_A^{-1} \vect{r}' \right\},
\end{equation}
with
\begin{align}
    \vect{r}' &= \matr{A}_{\vth} \vect{d} - \vect{\mu}'(\vth) \\
    \matr{\Sigma}_A &= \matr{A}^T_{\vth} \matr{\Sigma}_{\vth} \matr{A}_{\vth}.
\end{align}
While the two forms (\ref{eq:gaussian-likelihood} and \ref{eq:general-gaussian-likelihood}) are mathematically equivalent, it is sometimes convenient to use the scaled form in order to make certain properties of $\vect{\mu}$ clear, as we will now discuss.

\subsection{Inference Method}
\label{sec:amlp}
The models we will encounter in this work generally have a large number of parameters. This is prohibitive for performing Bayesian inference via MCMC, due to the `curse of dimensionality'. 
Examples of inference methods that are applicable to high-dimensional data (under some conditions) are Gibbs sampling and Hamiltonian Monte Carlo (HMC). 
However, these techniques do not easily yield the Bayesian evidence, which is useful for comparing models, and especially for deciding on the relevant number of parameters to include in our smooth models. 

Instead, we adopt a technique in which some of the parameters of the model are \textit{pre-marginalized}. 
That is, we integrate the posterior distribution analytically for the \textit{linear} parameters, reducing the effective dimensionality for the sampler, which must only deal with the remaining non-linear parameters. 
This technique has been previously described in \citep[eg.][]{Lentati2017,Monsalve2018,Tauscher2021}, and we derive it for our purposes in App. \ref{app:marginalized-likelihood}. 

In short, the result is that in the context of a particular MCMC sample, we must sample only a set of \textit{non-linear} parameters, for which we solve for the maximum-likelihood (ML) of the remaining linear parameters. Letting $\hat{\vect{r}}$ be the residuals of the data to this conditional ML model, the posterior of the non-linear parameters is given by
\begin{equation}
    p_{\rm NL}( \vth_{\rm NL} | \vect{d}) \propto \sqrt{|\matr{\Sigma}^{-1}| | \matr{\Sigma}_{\rm L}|} \exp\left\{ - \frac{1}{2} \hat{\vect{r}}^T \matr{\Sigma}^{-1} \vect{\hat{r}} \right\},
\end{equation}
where $\matr{\Sigma}_{\rm L}$ is the covariance matrix of the linear sub-model. 
It is also possible to obtain samples from the posterior of the linear sub-model via sampling from a multivariate normal with mean $\hat{\vth}_{\rm L}$ and covariance $\matr{\Sigma}_{\rm L}$ \citep[cf.][]{Tauscher2021}.


\subsection{Sampling Method}
\label{sec:sampler}
For all of our Bayesian sampling in this paper we use the \textsc{polychord} nested-sampling code \citep{Handley2015,Handley2015a}. For sampling, we use $N_{\rm live}\approx 100 N_{\rm dim}$, where $N_{\rm dim}$ is the number of parameters sampled by the MCMC (i.e. not including linear parameters). 
Importantly, \textsc{polychord} is able to generate an estimate of the Bayesian evidence, which is useful for model comparison.

For all non-linear parameters in this work we employ uniform priors, i.e. any value of each parameter -- within certain bounds -- is equally likely \textit{a priori}. 

\section{A Probabilistic Calibration Model}
\label{sec:calibration-model}
Measurements of EDGES' antenna temperature, like all antennas, is accompanied by some multiplicative gain and additive noise\footnote{In principle, the gain may in fact be non-linear, but such a case is actively avoided and indications of such a state of affairs are flagged in our processing. Thus, we proceed with the assumption of linearity of the gains.}.
While various external features, such as the angular response of the antenna, affect the voltage induced on the antenna itself, in this section we are not concerned with these effects, but rather the gain applied to this voltage on the signal path between the antenna and the analog-to-digital converter. 
That is, the gain applied by the receiver system itself before writing the measurements to disk. 

The primary -- but not only -- component that induces these gains is the low-noise amplifier (LNA), whose purpose is to amplify the incoming voltages from the antenna in order that additive noise in the rest of the system does not overwhelm the desired signal.
Unfortunately, the value of this (complex-valued) receiver gain is not constant -- either with frequency or time. It is dependent on the ambient temperature, humidity and other factors. 
To overcome this limitation, EDGES uses the well-known technique of Dicke-switching \citep{Rogers2012} to perform gross calibration of the receiver gains. In this technique, measurements switch between the input (ostensibly from the antenna) to two different internal reference loads. In practice, this technique is not sufficient on its own for the high-precision required of 21\,cm experiments; the signal path for the receiver input versus that of the internal references loads is slightly different (having an additional switch), and thus has slightly different reflection/propagation characteristics. These are accounted for by the noise-wave formalism \citep{Meys1978}. 

In this section, we present this technique of Dicke-switching along with the noise-wave formalism following \citet{Monsalve2017} (hereafter \msv). 
However, in doing so, we pay close attention to the probabilistic model, ultimately deriving a likelihood for the calibration parameters in a similar fashion to the recent work of \citet{Roque2020} for REACH. 
However, we do not follow \citet{Roque2020} in using conjugate priors to define our posterior distribution, using instead the linear marginalisation technique outlined in \S\ref{sec:amlp}.

All the quantities described in this section are frequency-dependent. 
However, to a good approximation, frequency channels are statistically uncorrelated in the observed spectra, and thus in this section we may consider each channel independently. Thus for notational clarity we omit frequency dependence throughout.

\begin{figure*}
    \centering
    \includegraphics[height=0.85\textheight,width=0.8\textwidth,keepaspectratio]{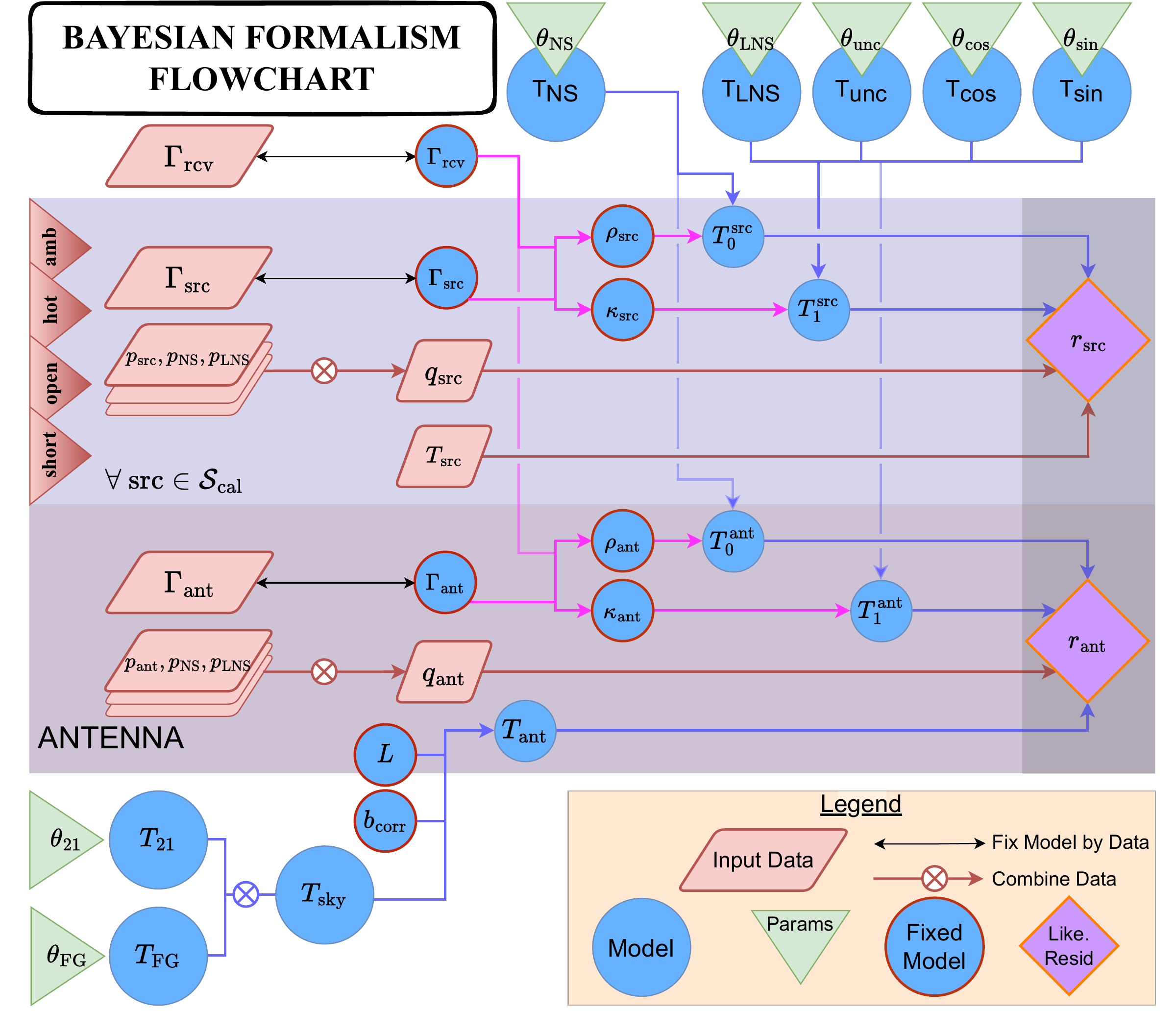}
    \caption{Flowchart for the Bayesian pipeline presented in this paper. The top shaded panel presents the pure calibration model (cf. \S\ref{sec:calibration-model}). By using the point estimates of the calibration parameters (green triangles at the top) along with the network in the lower shaded panel, one is executing the \textit{isolated} sky model fit (cf. \S\ref{sec:field-data-model}), while if all parameters are estimated together, the \textit{joint} inference is being performed. The top panel implicitly includes identical copies for each of the calibration sources, $\mathcal{S}_{\rm cal}$. Dark red lines follow the flow of spectrum and thermistor measurements. Pink lines trace the flow of reflection parameter measurements/models. Blue lines trace the flow of noise-wave and internal load temperature models. The likelihood is computed as a zero-mean multivariate Gaussian distribution with diagonal covariance, evaluated at the concatenation of all purple diamonds with orange outlines (i.e. $r_{\rm src}$). Blue circles with red outlines are models which are considered `fixed' by certain data in this analysis, though in principle in future analyses they should be left free. The only substantial difference between the calibration and antenna panels is that $T_{\rm ant}$ is a model instead of input data. Thus, the calibration parameters (top green triangles) are more effectively inferred from the calibration data, while the sky parameters (lower green triangles) require antenna data. In practice, the parameters co-vary and may be influenced indirectly through any measurement.}
    \label{fig:flowchart}
\end{figure*}

Since we introduce many variables throughout the next two sections, we provide a summary of the variable definitions in Tables \ref{tab:subscripts} and \ref{tab:variables}. The first provides a summary of the different sets of labels used throughout the paper, and the second lists many of the important variables.
Furthermore, we summarize the entire pipeline as a flowchart in Fig. \ref{fig:flowchart}.

\begin{table*}
    \centering
    \begin{tabularx}{\linewidth}{c|lp{2.2cm}lXl}
         \hline
         \textsc{Symbol} &  \textsc{Elements} & \textsc{Subsets} & \textsc{Vars} & \textsc{Description} & \textsc{Eqs.}\\
         \hline
         $\mathcal{L}$ & \{src, L, LNS\} & & switch & Internal switch-position for the receiver: `src' referring to receiver input port (may be substituted by label for the particular input source, see below), `L' to internal load, and `LNS' the internal `load plus noise-source' & \ref{eq:pswitch} \\
         \hline
         $\mathcal{S}$ & \{amb, hot, open, short, ant\} & $\mathcal{S}_{\rm cal} = \mathcal{S} - \{{\rm ant}\}$ & src & Sources attached to the receiver input port & \ref{eq:psrc} \\
         \hline
         $\mathcal{T}$ & \{unc, cos, sin, L, NS\} & {$\!\begin{aligned} \mathcal{T}_{\rm intload} &= \{{\rm L, NS}\} \\ \mathcal{T}_{\rm NW} &= \mathcal{T} - \mathcal{T}_{\rm intload} \\ \mathcal{T}_{\rm nl} &= \{{\rm NS}\} \\ \mathcal{T}_{\rm lin} &= \mathcal{T} - \mathcal{T}_{\rm nl}\end{aligned}$} & 
         $p$ & The five modeled temperature models (noise-waves and internal loads) & \ref{eq:tpoly} \\
    \hline
    \end{tabularx}
    \caption{Sets of labels used throughout this paper, with their defining symbols, and relevant subsets. The \textsc{Vars} column gives common variables used to stand in for elements of the set (usually as subscripts).}
    \label{tab:subscripts}
\end{table*}

\begin{table*}
    \centering
    \begin{tabularx}{\linewidth}{c|cXll}
         \hline
         \textsc{Symbol} &  & \textsc{Description} & \textsc{Domain} & \textsc{Eqs.} \\
         \hline
         $p_{\rm switch}$          & \meas       & Power from the receiver pointing to a given switch. & $\mathbb{R}^+$ & \ref{eq:pswitch} \\
         $q_{\rm src}$             & \meas       & The `three-position-switch ratio' which normalises input source power by measured internal powers & $\mathbb{R}$ & \ref{eq:basic-q}, \ref{eq:full-q}  \\
         $\vect{T}_{\rm NW}$       & \model      & Vector of noise-wave temperatures, $\left[T_{\rm unc}, T_{\rm cos}, T_{\rm sin}\right]$ at one frequency, specific to receiver. & $\mathbb{R}^3$ & \ref{eq:psrc} \\
         $T_{\rm src}$             & \meas|\model & Temperature of a source connected to the receiver input. Modeled for src=ant, measured otherwise. & $\mathbb{R}^+$ & \ref{eq:psrc} \\
         $\Gamma_{\rm inst}$, $\Gamma_{\rm src}$ & \fixedmodel & Reflection coefficients of the instrument and input sources respectively & $\mathbb{C}$, $0 \leq |\Gamma| \leq 1$ & \ref{eq:fsrc} \\
         $\vect{k}_{\rm src}$      & \fixedmodel       & 3-vector denoting the power transfer efficiency of an input source coupled to the receiver with respect to the noise-wave temperatures, \{unc, cos sin\} & $\mathbb{R}^3$ & \ref{eq:ksrc} \\
         $c_{\rm src}$             & \fixedmodel       & Power transfer efficiency of input source coupled to receiver with respect to input temperature & $\mathbb{R}$, $0 \leq c_{\rm src} \leq 1$ & \ref{eq:csrc} \\
         $h$                       & \fixedmodel       & Power transfer efficiency of the instrument & $ \mathbb{R}$, $0 < h < 1$ & \ref{eq:h} \\
         $\sigma^2_{q, {\rm src}}$ & \meas       & Variance of $q_{\rm src}$. Estimated empirically using time-samples as independent realizations for ${\rm src} \in \mathcal{S}_{\rm cal}$, and using residuals to high-order smooth polynomial fits over frequency for src=ant & $\mathbb{R}^+$ & \ref{eq:sigma2qsrc} \\
         $T'_{\rm L}$, $T'_{\rm NS}$ & \model    & Effective $(T_{\rm L}, T_{\rm NS})$ accounting for path differences between the source input and internal loads. & $\mathbb{R}^+$ & \ref{eq:per-source-linear} \\
         $\vect{T}$                & \model      & The vector of four temperatures that compose the linear sub-model: $[T_{\rm unc}, T_{\rm cos}, T_{\rm sin}, T'_{\rm L}]^T$ & $\mathbb{R}^{+4}$ & \ref{eq:per-source-linear} \\
         $T^{\rm src}_0$, $T^{\rm src}_1$ & \model & Multiplicative and additive temperatures converting measured $q_{\rm src}$ into source temperature, $T_{\rm src}$ & $\mathbb{R}^+$, $\mathbb{R}$ & \ref{eq:q2t}, \ref{eq:t0t1} \\
         $\vect{T}_p$              & \model      & Length-$N_\nu$ model temperature spectrum of one of the five estimated temperature models, $p \in \mathcal{T}$ & $\mathbb{R}^{N_\nu}$ & \ref{eq:tpoly} \\
         $\vect{\theta}_p$         & \param      & Length-$N^p_{\rm terms}$ vector of polynomial parameters for $\vect{T}_p$ & $\mathbb{R}^{N^p_{\rm terms}}$ &  \ref{eq:tpoly} \\
         $\matr{\Psi}$             & \choice      & $N_\nu \times N_{\rm terms}$ matrix of polynomial basis vectors, $\Psi_{ij} = (\nu_i/\nu_{\rm ref})^j$ & $\left(\mathbb{R}^+\right)^{N_\nu \times N_{\rm terms}}$ & \ref{eq:tpoly} \\
         $\vect{\theta}_{\rm NW+L}$& \param      & The vector of all polynomial coefficients for $p \in \mathcal{T}_{\rm lin}$: $[\vth_{\rm unc}, \vth_{\rm cos}, \vth_{\rm sin}, \vth_{\rm L}]$ & $\mathbb{R}^{N_{\rm NW+L}}$ & \ref{eq:dcal_gaussian} \\
         $\vect{r}_{\rm src}$& \both       & Model residual vector for ${\rm src} \in \mathcal{S}$ & $\mathbb{R}^{N_{\rm src} N_\nu}$ &  \ref{eq:rsrc} \\
         $\matr{\Sigma}_{\rm src}$ & \both       & The modeled diagonal covariance of $\vect{r}_{\rm src}$, equal to $T_{\rm NS}^2 \sigma^2_{q, {\rm src}}$ & $\left(\mathbb{R}^+\right)^{N_\nu}$ & \ref{eq:sigmarsrc}\\
         $c_{\rm terms}, w_{\rm terms}$ & \choice & Short-hand for the number of terms, $N^p_{\rm terms}$ used for $p\in\mathcal{T}_{\rm intload}$ and $p\in\mathcal{T}_{\rm NW}$ respectively & $\mathbb{Z}^+$ &  \\
         \hline
         $T_{\rm sky}$             & \model      & True radio temperature of the sky & $\mathbb{R}^+$ & \ref{eq:sky_temp} \\
         $T_{\rm 21}$              & \model      & Temperature of the cosmic 21\,cm radiation & $\mathbb{R}$ & \ref{eq:flattened-gaussian}\\
         $\overline{\vect{T}}_{\rm BWFG}$    & \model & LST-averaged beam-weighted foregrounds. Modeled as a linear sum of log-polynomials. & $\mathbb{R}^+$ & \ref{eq:linlog} \\
         $B$                       & \fixedmodel      & Antenna beam as a function of line-of-sight & $\mathbb{R}^+$ & \ref{eq:sky_temp_with_beam}\\
         $T_{\rm sky, beam}$       & \model      & Sky temperature after attenuation by antenna beam & $\mathbb{R}^+$ & \ref{eq:sky_temp_with_beam}\\
         $b_{\rm corr}$            & \fixedmodel      & Beam chromaticity correction & $\mathbb{R}$ & \ref{eq:beamcorr} \\
         $T_{\rm sky, bc}$         & \model      & Sky temperature after beam attenuation, but correcting for chromatic beam structure via $b_{\rm corr}$ & $\mathbb{R}^+$ & \ref{eq:sky_temp_bcorr} \\
         $L$                       & \fixedmodel      & Fractional loss in the signal path (includes antenna, balun, connector and ground loss) & $\mathbb{R}^+$, $0 < L < 1$ & \ref{eq:tsky_loss}, \ref{eq:loss}\\
         $p_{\rm sky, meas}$       &  \meas      & The measured power from the deployed antenna integrated over 39\,sec (uncalibrated) & $\mathbb{R}^+$ & \ref{eq:fully_uncalibrated_temp}\\
         $\widehat{\bar{T}}_{\rm sky, bc}$& \both& The estimated calibrated sky temperature, where calibration is derived from an iterative procedure, averaged over time. Equivalent to publicly available data. & $\mathbb{R}^+$ &  \ref{eq:processed_q} \\
         $\widehat{\bar{T}}'_{\rm sky, bc}$&\both& An estimate of the sky temperature obtained from decalibrating $\widehat{\bar{T}}_{\rm sky, bc}$ back to $q_{\rm ant}$ then re-calibrating with an alternate calibration model & $\mathbb{R}^+$ & \ref{eq:recalibrate}  \\
         \hline
    \end{tabularx}
    \caption{Summary of symbols used in this work, with a description, appropriate domain and example of equations in which they are used. The colored symbol in the second column provides a `type' for the quantity. \textbf{Key to symbols:} \model: `Models', \param: `Model Parameters',  \fixedmodel: `Fixed Models', \choice: `Model Choices' (eg. number of parameters), \meas: `Measurements', \both: `Model-transformed Measurements'. Note that not all listed model parameters or models are independent: some are derived from others, or simply concatenations of others. Note also that the distinction between `model' and `measurement' is not always simple. Here, by `measurement' we mean any symbol denoting a quantity that \textit{can} be directly measured, without requiring use of a model parameter. For example, $q_{\rm ant}$ can be calculated directly from measured data without requiring a model parameter. Conversely, `models' are here defined as quantities that, once model parameters are chosen, do not require any data to calculate and \textit{cannot} be uniquely determined by measurements. Notably, `measurements' as here defined may `modeled' (which is the point of inference), but are nonetheless defined as measurements under the above definitions. In-text equations involving symbols here defined as `measurements' may in fact be referring to either measurements or models, depending on context. `Fixed models' are those for which we do not let the parameters vary \textit{in this work}, but are also not direct measurements. `Model-transformed measurements' are those quantities for which both the measured data \textit{and} some choice of model parameter is required in order to calculate.}
    \label{tab:variables}
\end{table*}

\subsection{The Noise-Wave Formalism}
\label{sec:model:threepos}


The Dicke switching technique in EDGES alternates between three switch positions: the input `source' (src; typically the antenna), an internal `load' (L) and an internal `load + noise-source' (LNS).
Given a `true' temperature, $T_{\rm switch} \in \{T_{\rm src}, T_{\rm L}, T_{\rm LNS}\}$, for any of these switches at a particular frequency, the receiver imparts a time-dependent multiplicative gain, and adds its own noise, such that the output power is
\begin{equation}
    p_{\rm switch} = g T_{\rm switch} + T_{\rm inst} + n_{\rm switch},
    \label{eq:pswitch}
\end{equation}
where $T_{\rm inst}$ is the instrument's thermal contribution, and $n_{\rm switch}$ is a zero-mean Gaussian random variable whose variance is proportional to $g T_{\rm switch}$\footnote{We do not provide a definite form for $n_{\rm switch}$ here, as its true form is dependent on a number of subtle factors, such as the spectrometer and internal noise characteristics. For this paper, it is enough to assume it is zero-mean and Gaussian.}.

We may thus form the power quotient
\begin{align}
    q_{\rm src} &\equiv \frac{p_{\rm src} - p_{\rm L}}{p_{\rm LNS} - p_{\rm L}},
    \label{eq:basic-q}
\end{align}
We note that the numerator and denominator are both Gaussian-distributed, and are correlated due to their mutual dependence on the realization of the load power, $p_{\rm L}$. 
Here, $q_{\rm src}$ is a random value for a \textit{single integration} (i.e. approximately 40 seconds worth of total measurement). We shall denote an average of $N$ such integrations as $\bar{q}_{\rm src}$.
Note that to first-order, the receiver gain is cancelled in $q_{\rm src}$, as it is present in each of the terms in both numerator and denominator\footnote{This assumes, of course, that the receiver gain is stable over timescales of $\sim40\,{\rm sec}$.}.

The three measured powers may be modeled using the noise-wave formalism \citep{Meys1978}. Following \msv, we write 
\begin{align}
    \langle p_{\rm src} \rangle &= g \left[c_{\rm src} T_{\rm src} + \vect{k}_{\rm src}\cdot \vect{T}_{\rm NW} \right] + T_{\rm inst}, \label{eq:psrc} \\
    \langle p_{\rm L} \rangle  &= g^\star \left[h T_{\rm L}\right] + T_{\rm inst} , \label{eq:pload}\\
    \langle p_{\rm LNS} \rangle &= g^\star \left[h (T_{\rm L} + T_{\rm NS})\right] + T_{\rm inst}, \label{eq:plns}
\end{align}
where
\begin{equation}
    h = 1 - |\Gamma_{\rm inst}|^2
    \label{eq:h}
\end{equation}
is the `power transfer efficiency' of the instrument and $\Gamma_{\rm inst}$ is the complex-valued reflection coefficient of the receiver (measured in terms of the `$S_{11}$'). 
Here, $\vect{T}_{\rm NW} = [T_{\rm unc}, T_{\rm cos}, T_{\rm sin}]^T$ are the \textit{noise-wave} temperatures, which quantify standing-wave contributions of the noise reflected from the receiver back to the antenna. Here `unc' refers to the uncorrelated portion of the noise-wave, while `cos' and `sin' refer to the two correlated portions that in and out of phase.

Further note that the frequency-dependent gain is different for the input source and the internal loads. This is due to a small internal path difference on account of the switch. It is convenient to write $g^\star = g(1 + \delta_g)$. 

The coefficients $\vect{k}_{\rm src}$ and $c_{\rm src}$ are frequency-dependent functions of the reflections coefficients of the instrument, $\Gamma_{\rm inst}$ and input source, $\Gamma_{\rm src}$.
\msv\ provides the values of the coefficients as
\begin{align}
    c_{\rm src} &= \left(1 - \left|\Gamma_{\rm src}\right|\right)^2 F_{\rm src}^2 \label{eq:csrc}\\
    \vect{k}_{\rm src} &= \left[|\Gamma_{\rm src}|^2 |F_{\rm src}|^2, \ 
        |\Gamma_{\rm src}| |F_{\rm src}| \cos \alpha_{\rm src},\ 
        |\Gamma_{\rm src}| |F_{\rm src}| \sin \alpha_{\rm src} \right]
        \label{eq:ksrc}
\end{align}
where
\begin{align}
    \ \ 
    F_{\rm src} = \frac{\sqrt{h}}{1 - \Gamma_{\rm inst}\Gamma_{\rm src}}, \ {\rm and}\ 
    \alpha_{\rm src} = {\rm arg}(\Gamma_{\rm src}F_{\rm src}).
    \label{eq:fsrc}
\end{align}
Both $\Gamma_{\rm src}$ and $\Gamma_{\rm inst}$ are independently measured in the lab (or, in the case of the antenna, repeatedly in the field), with their own thermal and systematic uncertainties. 
In general, our calibration likelihood should directly include the raw $S_{11}$ measurements (from which $\Gamma_{\rm inst}$ and $\Gamma_{\rm src}$ are computed), with an estimate of their noise properties. 
However, our focus in this paper is not the estimation of $\Gamma$, and we ignore the thermal uncertainty in the measurements for now, instead using best-fit Fourier-series models to characterize $\Gamma_{\rm src}(\nu)$ and $\Gamma_{\rm inst}(\nu)$. 

\begin{figure}
    \centering
    \includegraphics[width=\linewidth]{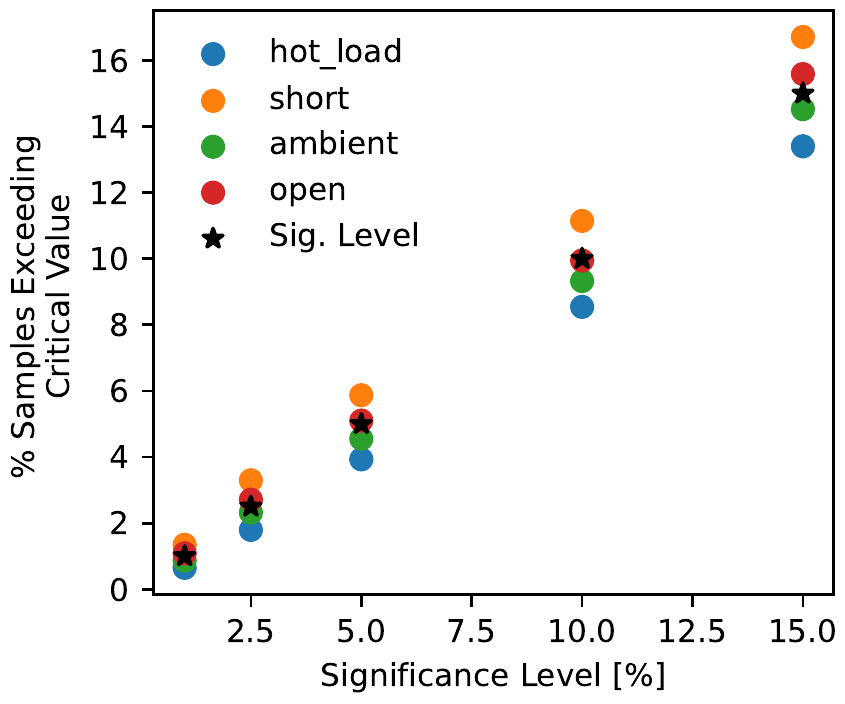}
    \caption{Validation that the distribution of \textbf{Q} is Gaussian. The Anderson-Darling (AD) metric was computed for all samples in a particular channel (where samples were taken over time from a single calibration input). This plot shows the percentage of channels where the AD metric was greater than the critical value for a particular significance level, i.e. the number of channels for which \textit{rejection} of the hypothesis of Gaussianity is expected to be inappropriate to a certain level. 
    For example, the right-most points show the percentage of channels for which Gaussianity can be rejected, while expecting 15\% of the rejections to be incorrect. Colored points above each black star indicate that there is an excess of number of channels for which Gaussianity can be rejected, providing some evidence that the total spectrum has some non-Gaussianity. We find that only the shorted cable has some evidence of non-Gaussianity, and it is marginal (at most an excess of $\sim2\%$ of channels are considered non-Gaussian).
    }
\label{fig:gaussianity}
\end{figure}

Inserting the models for the internal powers into \autoref{eq:basic-q}, we find
\begin{equation}
    q_{\rm src} =  \frac{c_{\rm src} T_{\rm src} + \vect{k}_{\rm src}\cdot \vect{T}_{\rm NW} -h T_{\rm L} + \left[n_{\rm src} - n_{\rm L}\right]/g}{(1+\delta_g) h T_{\rm NS} + \left[n_{\rm LNS} - n_{\rm L}\right]/g}.
    \label{eq:full-q}
\end{equation}
While the distribution of the noise in the numerator and denominator are both Gaussian, the distribution of $q_{\rm src}$ is not in detail -- it is the distribution of a ratio of  correlated Gaussian random variables. 
In general, with knowledge of the variance of $(n_{\rm src}, n_{\rm L}, n_{\rm LNS})$, which themselves depend on the various temperatures involved, one can derive the distribution of $q_{\rm src}$. 
In practice, doing so is rather complicated, and we defer this computation to future work. 
In this paper, we merely note that if $T_{\rm NS}$ is large compared to 
$T_{\rm L}$ (which is true for EDGES), the distribution of $q_{\rm src}$ is empirically close to Gaussian (cf. Fig. \ref{fig:gaussianity}), with some evidence in our particular data for small non-Gaussianities in our shorted cable input \citep[for models that account for non-Gaussianities, see][]{Scheutwinkel2022a}. 
We may also approximate the covariance as diagonal, as long as we average together $\sim$16 adjacent raw frequency channels \citep[cf.][for details]{Murray2022}.
It can thus be approximately described simply by its expectation, 
$\langle q_{\rm src} \rangle$ and variance, $\sigma^2_{q,{\rm src}}$. That is, we approximate
\begin{align}
    q_{\rm src} &\approx \langle q_{\rm src} \rangle + n_{\rm src}  \label{eq:qsrc-approx}\\
    n_{\rm src} &\sim \mathcal{N}\left(0, \sigma^2_{q, {\rm src}}\right). \label{eq:nsrc}
\end{align}

To estimate the variance, we assume the time axis to be statistically stationary\footnote{This has been verified for calibration sources by using an augmented Dickey-Fuller test, which yields $p$-values of order $10^{-28}$ or less for all sources. Note that this assumption is only made for \textit{calibration sources}, not the in-field antenna, whose variance has siderial dependence.} so that we compute
\begin{equation}
    \sigma^2_{q, {\rm src}} = \langle (q_{\rm src}(t) - \bar{q})(q_{\rm src}(t) - \bar{q}) \rangle_t.
    \label{eq:sigma2qsrc}
\end{equation}

The expectation, $\langle q_{\rm src} \rangle$, can be approximated by taking the second-order Taylor expansion of the expectation of a ratio, \autoref{eq:stat:expectation-ratio}, applying it to the RHS of \autoref{eq:basic-q}.
We find that 
\begin{equation}
    \langle q_{\rm src} \rangle \approx \frac{c_{\rm src} T_{\rm src} + \vect{k}_{\rm src}\cdot \vect{T}_{\rm NW} -h T_{\rm L}}{(1+\delta_g) h T_{\rm NS}} \left(1 - \delta_0 + \delta_1^2 - \cdot\right),
    \label{eq:expectation-q}
\end{equation}
where the $\delta_i$'s are small dimensionless numbers dependent on the various source 
temperatures\footnote{In detail, this expansion makes $q_{\rm src}$ \textit{not} linear 
in $T_{\rm src}$, as it appears in the $\delta_i$ terms in complicated ways. Nevertheless, 
this effect is small so long as $T_{\rm NS}$ is large.}. We ignore the terms involving
$\delta_i$ in this work, as they are very small (as long as $T_{\rm NS}$ is sampled with high signal-to-noise, and is large compared to $T_{\rm L}$).

We would like to solve for the noise-wave parameters and the internal temperatures of the receiver. 
Notice that with the exception of $T_{\rm NS}$, the equation is linear in 
its parameters. This is made more clear by re-writing our model as
\begin{equation}
    T'_{\rm NS} \langle q_{\rm src} \rangle  - \rho_{\rm src} T_{\rm src} \approx \frac{\vect{k}_{\rm src}}{h} \cdot \vect{T}_{\rm NW}- T'_{\rm L} = \vect{\kappa}_{\rm src} \cdot \vect{T}
    \label{eq:per-source-linear}
\end{equation}
where 
\begin{align}
    \rho_{\rm src} &= c_{\rm src}/h, \\
    T'_{\rm L} &= (1 + \delta_g)T_L \\
    T'_{\rm NS} &= (1 + \delta_g)T_{\rm NS}, \\
    \vect{\kappa}_{\rm src} &= [\vect{k}_{\rm src}/h, -1] \ \ {\rm and} \\
    \vect{T} &= [\vect{T}_{\rm NW}, T'_{\rm L}].
\end{align}
Since $T'_{\rm L}$ and $T'_{\rm NS}$ share the same essential properties as $T_{\rm L}$
and $T_{\rm NS}$ (i.e. they are smooth over frequency), it is just as reasonable to 
estimate them instead. 

Inverting Eq.\ref{eq:per-source-linear}, we find that an estimate of the input source temperature may equivalently be written as a linear transformation of the measured $q_{\rm src}$:
\begin{equation}
    \widehat{T}_{\rm src} = T^{\rm src}_0 q_{\rm src} + T^{\rm src}_1,
    \label{eq:q2t}
\end{equation}
where the sampling distribution of $\widehat{T}_{\rm src}$ is Gaussian with variance $T_0^{\rm src} \sigma^2_{\rm q, src}$, and 
\begin{equation}
    T^{\rm src}_0 = T'_{\rm NS} / \rho_{\rm src}, \ \ \ {\rm and}\ \ \ 
    T^{\rm src}_1 =  - \vect{\kappa}_{\rm src} \cdot \vect{T} / \rho_{\rm src}.
    \label{eq:t0t1}
\end{equation}
These two temperatures will be helpful in understanding the overall multiplicative and additive effects of the signal chain.

\subsection{A Naive Calibration Likelihood}
\label{sec:cal:naive}
To infer the noise-wave parameters, the EDGES experiment takes the receiver to the lab, and replaces the antenna with four \textit{known} input sources,  ${\rm src} \in \mathcal{S}_{\rm cal}$, where $\mathcal{S}_{\rm cal}$ is the set $\{ {\rm amb, hot, short, open} \}$.
Each source in $\mathcal{S}_{\rm cal}$ has different reflection characteristics as a function of frequency. 

We measure three primary quantities for each source as a function of frequency: (i) spectra, $\vect{q}_{\rm src}$, (ii) physical temperature, $\vect{T}_{\rm src}$ and (iii) reflection coefficient $\vect{\Gamma}_{\rm src}$. Of these, in this paper we consider only the spectra to have non-negligible uncertainty.

We seek to generate posteriors on models for the noise-wave temperatures as well as the load and noise-source temperatures.
We introduce some book-keeping notation for these sets of parameters; let $\mathcal{T}_{\rm NW}$ be the set of labels corresponding to the noise-wave terms: $\mathcal{T}_{\rm NW} = \{{\rm unc, cos, sin}\}$, and $\mathcal{T}_{\rm intload} = \{ {\rm L, NS}\}$ the labels corresponding to internal load temperature terms. Then the full set of modeled temperature terms is $\mathcal{T} = \mathcal{T}_{\rm NW} \cup \mathcal{T}_{\rm intload}$. An alternative useful partition of $\mathcal{T}$ is into the terms that can be treated as linear (in the sense of App. \ref{app:marginalized-likelihood}), $\mathcal{T}_{\rm lin} = \mathcal{T}_{\rm NW} \cup \{ {\rm L}\}$ and those that must be considered non-linear, $\mathcal{T}_{\rm nl} = \{ {\rm NS} \}$. 

The modeled temperature noise-wave temperatures and the load and noise-source temperatures are not arbitrary;
they are assumed to be smooth functions of frequency. 
We thus model each temperature as a low-order polynomial:
\begin{equation}
    \vect{T}_p = \sum_i^{N^p_{\rm terms}} \vth^p_i \left(\frac{\vect{\nu}}{\nu_0}\right)^i = \matr{\Psi} \vth_p, \ \ p \in \mathcal{T}
    \label{eq:tpoly}
\end{equation}
where $\vect{\nu}$ is the vector of observed frequencies (and the exponentiation is implicitly element-wise), $\vth_p$ are the unknown coefficients for temperature $p$ (in temperature units) and $\matr{\Psi}$ is the $N_\nu \times N^p_{\rm terms}$ matrix of polynomial basis vectors.

Let $\vect{r}_{\rm src}$ be the length-$N_\nu$ model-residual vector for a particular input source, ${\rm src} \in \mathcal{S}_{\rm cal}$:
\begin{equation}
    \vect{r}_{\rm src} = \vect{q}_{\rm src} \circ \vect{T}_{\rm NS} - \vect{\rho}_{\rm src} \circ \vect{T}_{\rm src} - \sum_{p \in \mathcal{T}_{\rm lin}} \vect{\kappa}_{p, {\rm src}} \vect{T}_p.
    \label{eq:rsrc}
\end{equation}
Under our assumptions of Gaussianity of $\vect{q}$, and independence between frequency channels, as justified in the previous subsection, we then have that the distribution of $\vect{r}_{\rm src}$ is a multivariate Gaussian with zero mean and diagonal covariance given by
\begin{equation}
    \matr{\Sigma}_{\rm src} = \matr{I}_{N_\nu} \vect{\sigma}^2_{q,{\rm src}} \vect{T}_{\rm NS}^2.
    \label{eq:sigmarsrc}
\end{equation}
Then, our final calibration likelihood is
\begin{align}
   \mathcal{L}_{\rm cal}(\vth_{\mathcal{T}}) &= \prod_{{\rm src}\in \mathcal{S}_{\rm cal}} \mathcal{L}_{\rm src}(\vect{q}_{\rm src} | \vth_{\mathcal{T}}), \label{eq:cal_likelihood_conceptual}\\ 
   \mathcal{L}_{\rm src}(\vect{q}_{\rm src} | \vth_{\mathcal{T}}) &\propto  \sqrt{ \left|\widehat{\matr{\Sigma}}^{-1}_{\rm src}\right|} \exp\left\{-\frac{1}{2}\vect{r}^T_{\rm src} \widehat{\matr{\Sigma}}_{\rm src}^{-1}\vect{r}_{\rm src}\right\}, 
    \label{eq:src_likelihood_conceptual}
\end{align}

This is conceptually the simplest representation of the likelihood, but for the purpose of exploiting the analytic marginalization of linear parameters (cf. \S\ref{sec:amlp}), it is helpful to separate the linear parameters into a single term represented by a product of a matrix with the linear parameter vector. App. \ref{app:linear-cal-model} details this process, showing that $\vth_p$ are linear for $p \in \mathcal{T}_{\rm lin}$.

\subsection{Comparison of likelihood to iterative approach}
\label{sec:differences-iterative}

The fiducial calibration temperatures, $\vect{T}_p$, used in \np\ were determined by an iterative process outlined in \msv. 
This process has several differences with respect to the calibration likelihood presented here, which result in somewhat differing calibration solutions.
Our purpose in this paper is to understand the \textit{posterior distribution} of the inferred cosmic signal, given uncertainties in the calibration parameters. 
To do this, we wish to keep the `point estimate' of the calibration solutions rather similar to the results of \np, by choosing methods and other parameters that match as closely as possible. 
Thus, it is important to understand where differences in the point estimates of the calibration arise with respect to the previous methods, before moving on to propagating uncertainties forward to field data.

The primary differences between the solutions used in \np\ and point estimates (nominally maximum-likelihood estimates) from our likelihood are as follows:
\begin{enumerate}
    \item \np\ smoothed calibration data into 8-channel bins (with a Gaussian filter) before fitting calibration polynomials, whereas we bin the spectra into 32-channel bins with a top-hat filter. The purpose of this change is to ensure that the covariance is diagonal (cf. \S\ref{sec:model:threepos}).
    \item \np\ inherently treated each frequency and source with the same weight (i.e. variance). Since doing so would bias our posterior distribution, we cannot use this assumption, and instead use the empirically-determined variance for each source and frequency.
    \item The iterative method separates sources and model parameters. The \texttt{amb} and \texttt{hot} sources are essentially zero-length cables with extremely good impedance match to the receiver. This means that $\Gamma_{\rm amb}$ and $\Gamma_{\rm hot}$ are extremely small. Conversely, the cable measurements (\texttt{open} and \texttt{short}) are designed to have high reflections, which makes it possible to characterize the noise-wave temperatures. Since the solutions are very sensitive to the accuracy of the $\Gamma_{\rm src}$ measurements, the iterative solutions for $\vect{T}_\mathcal{T}$ use the cable measurements \textit{only} to directly fit the noise-wave terms, $\mathcal{T}_{\mathcal{T}_{\rm NW}}$, using \texttt{amb} and \texttt{hot} to fit the internal load temperatures, $T_{\mathcal{T}_{\rm intload}}$. This avoids leaking any potential biases from inaccurate $\Gamma_{\rm open/short}$ measurements to $T_{\mathcal{T}_{\rm intload}}$. This is not possible for the likelihood, which consistently accounts for all data.
\end{enumerate}

\begin{figure*}
    \centering
    \includegraphics[width=\linewidth]{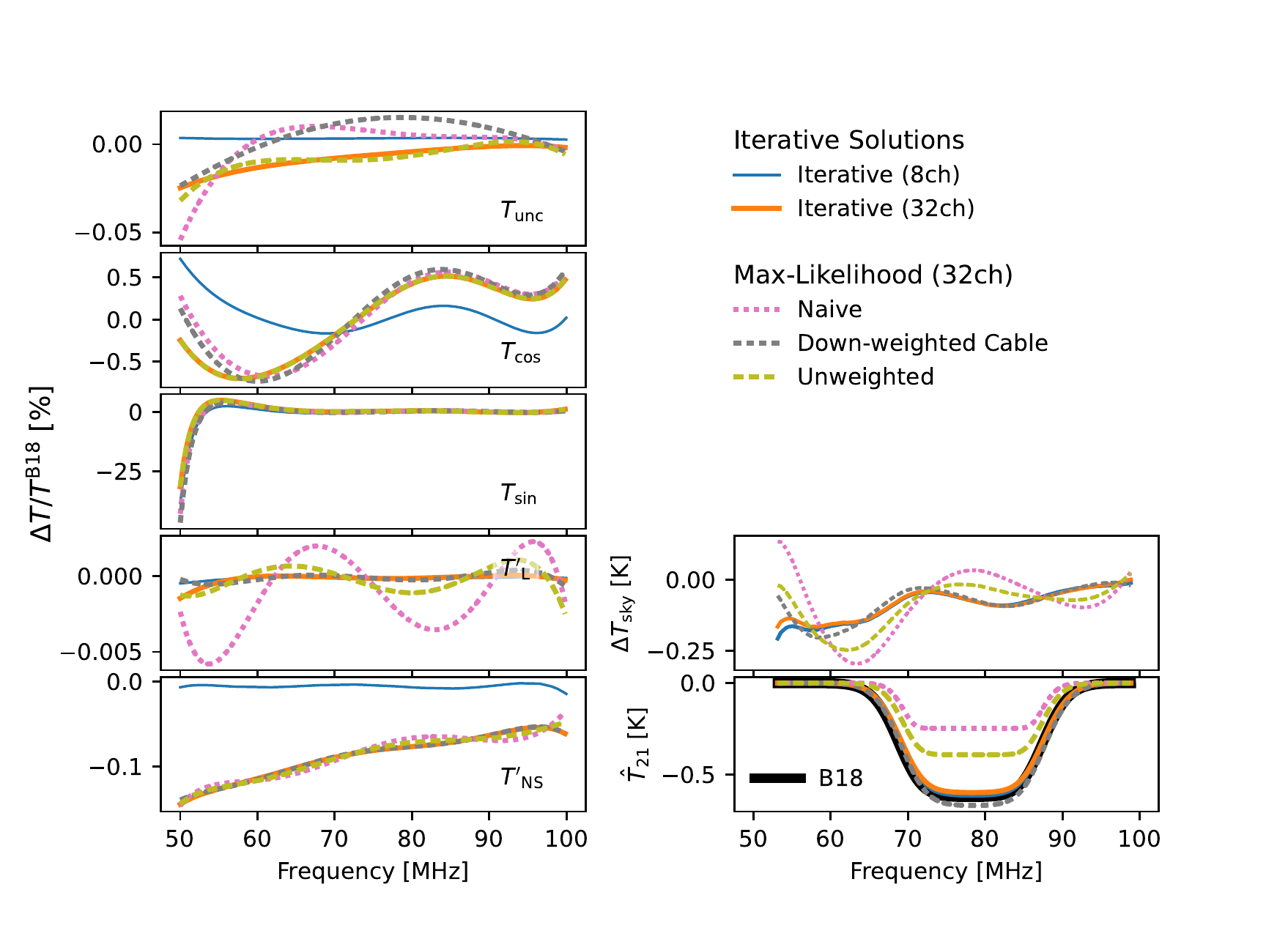}
    \caption{Summary of differences in calibration solutions with respect to those used in \np. The left column shows the five calibration parameters as a function of frequency, each displayed as a percentage deviation from \np. The right panels show the effect of the deviated solutions on the sky data, with the upper plot showing the absolute difference in (re)calibrated sky temperature (cf. Eq. \ref{eq:recalibrate}), and the lower plot showing the `best-fit' cosmic signal assuming a 5-term LinLog foreground model (Eq. \ref{eq:linlog}). Solid blue curves represent a calibration performed with the traditional iterative algorithm of \msv, and these match \np\ very well. Smoothing over 32 channels (solid orange) does not affect the inferred signal significantly. All other (dashed) curves represent maximum-likelihood calibrations using Eq. \ref{eq:cal_likelihood_conceptual}, with varying assumptions. Pink is our naive likelihood in which all sources and frequencies have variance empirically estimated. This results in large differences with respect to \np\ for the inferred signal. Assuming a constant variance with respect to frequency and sources (yellow) increases correspondence, but is still quite discrepant. The discrepancy can be largely removed by down-weighting the cable measurements (grey), minimizing their contribution to the estimation of the offset and scale parameters ($T'_{\rm L}$ and $T'_{\rm NS}$). See text for details.}
    \label{fig:differences_with_alan}
\end{figure*}

Fig. \ref{fig:differences_with_alan} summarizes the differences in the calibration solutions for our likelihood compared to those used in \np, where all choices are kept as similar as possible with the exception of the three differences just mentioned.
Note that \np\ uses $N^p_{\rm terms} = c_{\rm terms} = 6$ for $p \in \mathcal{T}_{\rm intload}$ and $N^p_{\rm terms} = w_{\rm terms} = 5$ for $p \in \mathcal{T}_{\rm NW}$. 
We plot each calibration solution individually in the left-hand panels, as a percentage difference from the solution used in \np.
In the top right-hand panel we plot the induced absolute difference in the calibrated, beam-corrected sky temperature:
\begin{equation}
    \Delta T_{21} = \hat{\bar{T}}^{\rm B18}_{\rm sky,bc} - \hat{\bar{T}}_{\rm sky,bc},
\end{equation}
where $\hat{\bar{T}}^{\rm B18}_{\rm sky,bc}$ is the publicly-available calibrated sky temperature from \np\, and $\hat{\bar{T}}_{\rm sky,bc}$ is obtained by \textit{de}-calibrating $\hat{\bar{T}}^{\rm B18}_{\rm sky,bc}$ with the exact \np\ calibration solution, and \textit{re}-calibrating with a new solution as specified in the legend (via Eq. \ref{eq:recalibrate}). 
The lower right-hand panel shows the inferred cosmic signal from this new `recalibrated' spectrum. Note that the inferred cosmic signals shown here are \textit{not} jointly estimated along with the calibration, as we shall do in \S\ref{sec:field-data-model}. 
Instead, after recalibrating the spectrum we fit a simple model, $T_{\rm sky} = T_{\rm FG} + T_{21}$, where the foreground term is given by a 5-term `linlog' model (cf. Eq. \ref{eq:linlog}) and $T_{21}$ is a flattened Gaussian (cf. Eq. \ref{eq:flattened-gaussian}).
The fit over the 21\,cm parameters is performed via global minimization routine, where for each parameter-set choice, we find the MLE of the foregrounds using linear algebra. 
Thus, this lower-right panel gives a fast indication of how much the difference in calibration affects our target inference.

We first concentrate attention on the solid blue line, representing the solutions using the same iterative procedure as used in \np\ (but with updated Python code instead of the original C-code). 
While some small differences are apparent (especially in $T_{\rm sin}$, which deviates by tens of percent at the lowest frequencies), their overall effect is extremely small, as evidenced by the excellent agreement of the inferred cosmic signal. 
To test difference (i) -- i.e. increase frequency bin size -- we plot in orange the result of binning in 32-channel bins, and still performing an iterative fit.
While this results in some marked differences in the calibration parameters (up to 1\% in $T_{\rm cos}$), the differences in the inferred cosmic signal are negligible.
The likely reason for the differences, especially in the scaling, $T'_{\rm NS}$, is the change from a Gaussian filter to a top-hat filter (blue to orange), rather than the change in bin size. 
We find that for a bin size of 32 channels (both for the top-hat and the Gaussian filter), the top-hat filter systematically produces higher average values for the hot load spectrum, at the level of $\sim0.05\%$.
A higher hot load temperature corresponds to a matching \textit{decrease} in $T'_{\rm NS}$, as witnessed. 
We choose the top-hat filter because it induces less correlation between neighbouring bins. 
Regardless of this difference, as already mentioned the estimated 21\,cm signal is essentially unaffected. 
This is for two reasons: (i) the differences are smooth in frequency and largely ameliorated by the flexible foreground model, and (ii) the calibration parameters have correlated effects which partially cancel each other in the final calibrated spectrum.

However, we find a very strong difference when we use our default maximum-likelihood model (pink dotted line). In particular, we note how the calibration solutions incur extra spectral structure, and the inferred cosmic signal is significantly more shallow. 
We first ask whether this difference may be due to difference (ii), i.e. the fact that our likelihood accounts for frequency- and source- dependent noise. 
The dashed yellow curves shows the results of a likelihood in which the variance is constant (since this figure show a maximum-likelihood estimate, the magnitude of the variance is inconsequential) over frequency and source. 
While the agreement between this curve and \np\ is somewhat better, this does not seem to be the dominant difference.

\begin{figure}
    \centering
    \includegraphics[width=\linewidth]{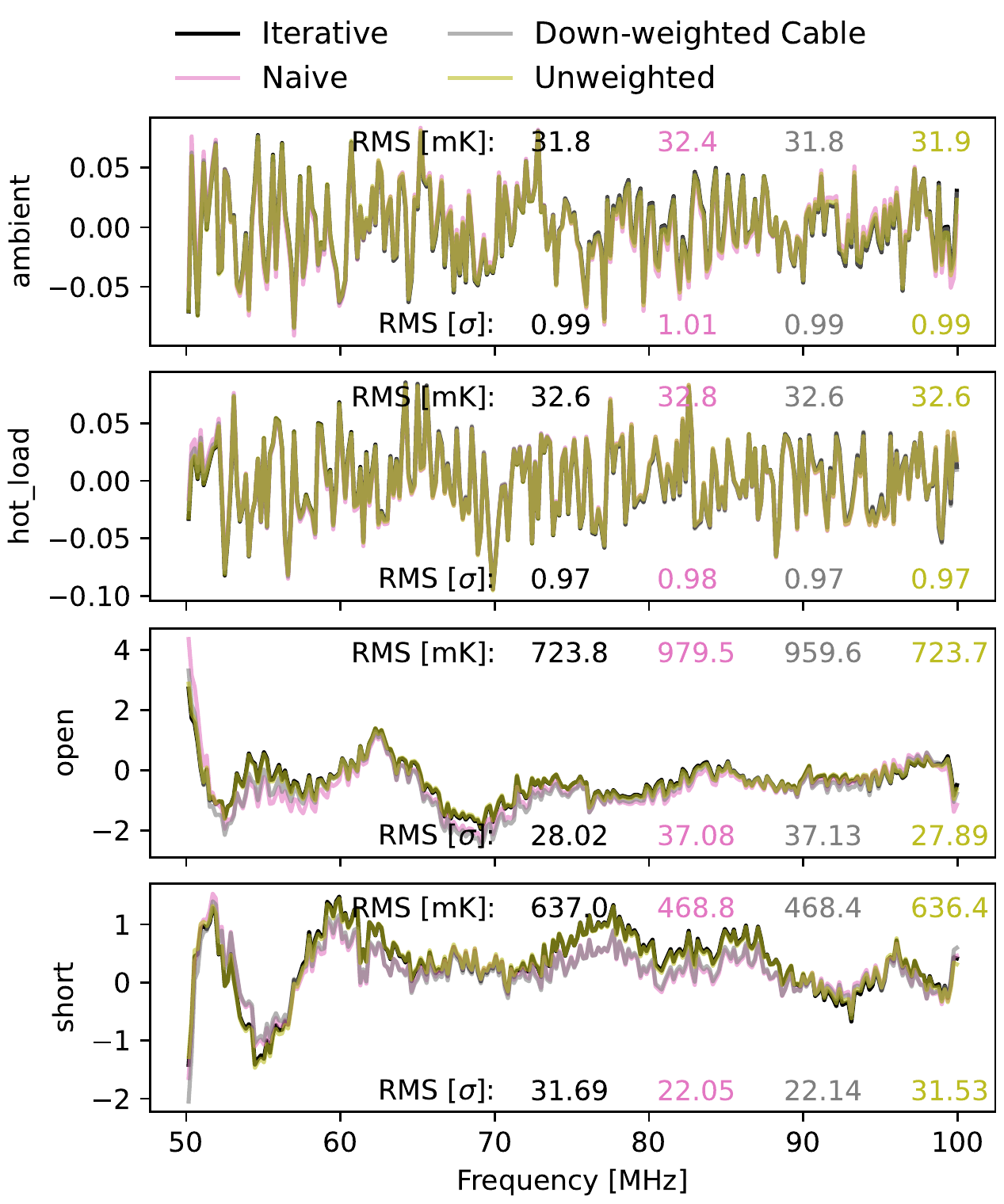}
    \caption{Residuals to the known temperature for each input source, for a range of likelihood models (see caption of Fig. \ref{fig:differences_with_alan} for a key to the models).  Also listed for each model and source are the RMS of the residuals, in units of mK and also numbers of standard deviation (according to the empirical noise model). Overall, differences between models are small. However, the cable measurements (bottom two rows) have larger weighted RMS (in terms of numbers of standard deviations) when their contribution is down-weighted. Conversely, when down-weighting the cable measurements, the fit to the ambient and hot load are significantly improved.}
    \label{fig:lab-resids}
\end{figure}

Difference (iii) concerns weighting different input sources for different parameters. 
The justification for the iterative method (largely) ignoring the contribution of the cable for estimating $T'_{\rm L}$ and $T'_{\rm NS}$ is that there are potential biases in the measurements of $\Gamma_{\rm open}$ and $\Gamma_{\rm short}$ that skew the estimates.
Fig. \ref{fig:lab-resids} demonstrates that for this calibration dataset, this is indeed the case. It shows significant (>20$\sigma$) `wiggles' in the \texttt{open} and \texttt{short} residuals for the iterative approach (and all approaches). 
The default likelihood (solid pink) is able to achieve far better precision on the \texttt{short} measurement, but sacrifices accuracy on all other measurements to do so.
Since the structure being fit in the \texttt{short} measurement is \textit{a priori} expected to be due to biases in $\Gamma_{\rm short}$, the increased residuals in the other measurements can be considered `leakage' from the bias in \texttt{short}. 
In other words, we would like the estimation of scale and offset parameters to be dominated by the very accurate \texttt{ambient} and \texttt{hot\_load} measurements, restricting any potential biases in the cable measurements to affect the noise-wave parameters.
To check whether this difference is dominating our discrepancy with \np, we fit a likelihood in which the cable measurements are significantly down-weighted, by artificially increasing their variance\footnote{We note that doing so is \textit{not self-consistent}. While we may achieve reasonable maximum a-posteriori estimates based on intuitive expectations with this approach, the posteriors will have an incorrect spread, and the Bayesian evidence will be wrong.}.
The result in both Figs. \ref{fig:differences_with_alan} and \ref{fig:lab-resids} is shown in grey. 
Importantly, the scale and offset parameters exhibit far less spectral structure than the default likelihood (pink dotted), and the inferred cosmic signal is in much better agreement with \np. In Fig. \ref{fig:lab-resids}, we notice that while the `down-weighted cable' model performs similarly to the default likelihood in calibrating the cable measurements, its errors there do not leak into the loads. 
The remaining differences between the down-weighted cable likelihood and \np\ are a combination of the frequency- and source-dependent weighting, and differences in the exact strength with which the cable measurements are down-weighted (the iterative method does not afford an obvious translation of its effective weighting of the measurements).

\subsection{A De-biased Calibration Likelihood}
\label{sec:results-pure-calibration}
The previous section showed that there is a significant systematic in the cable measurements used to derive our calibration, and that the effect of this systematic can be largely avoided by minimizing the impact of the cable measurements on the estimation of the load and noise-source temperatures. 
In this subsection, we outline a modified likelihood that takes advantage of this knowledge so as to decrease its bias.
We note that the `down-weighted cable' likelihood used in \S\ref{sec:differences-iterative} is not appropriate, since its noise model is known to be incorrect, and therefore it will yield incorrect posterior distributions and Bayesian evidence.

\begin{figure}
    \centering
    \includegraphics[width=\linewidth]{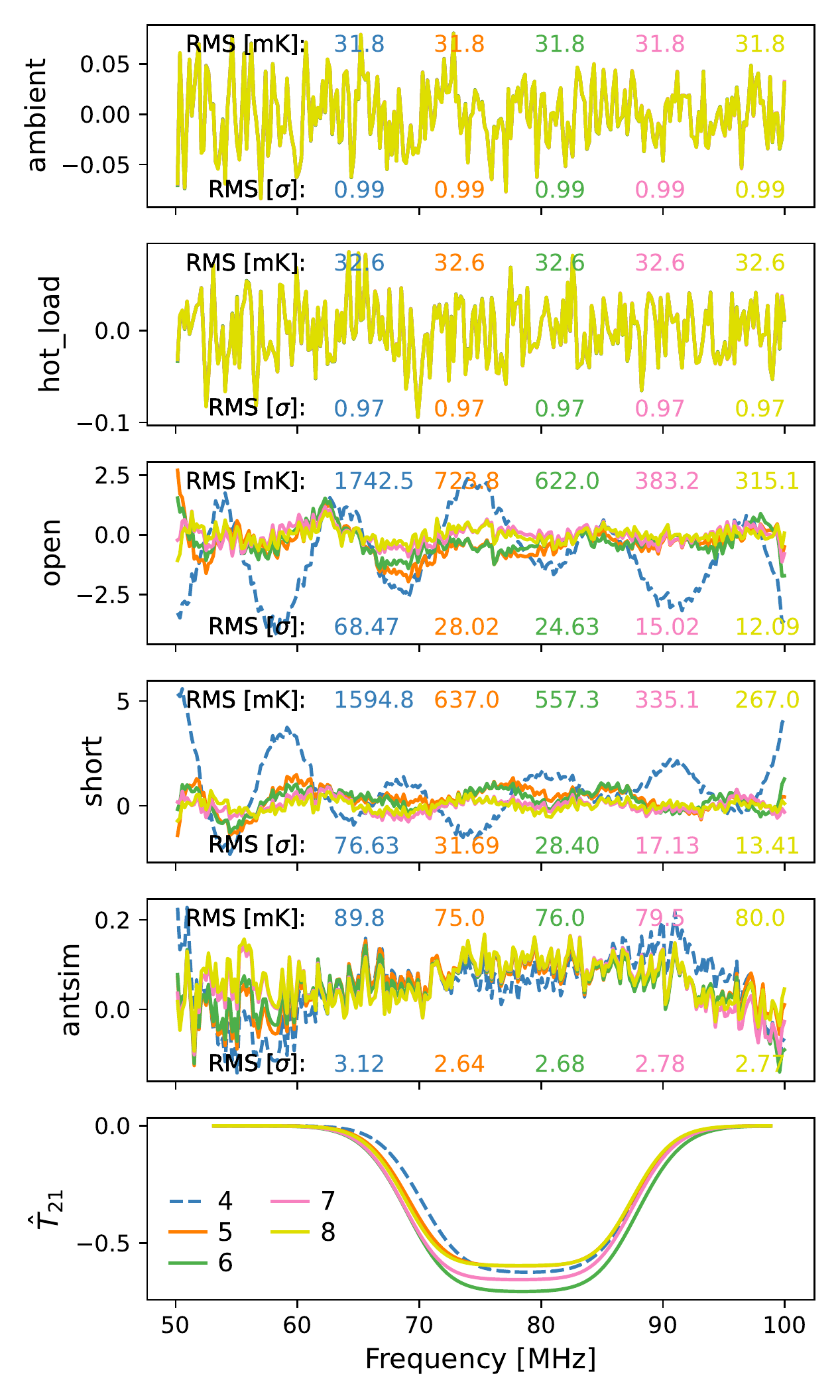}
    \caption{Calibration source residuals to models with different number of $w_{\rm terms}$. As $w_{\rm terms}$ increases, the load residuals (top two panels) stay the same, as they are not sensitive to $w_{\rm terms}$. The cable measurements (third and fourth panels) show ever-decreasing residuals, but with a mean $\chi^2 > 10$ even for the highest terms. We affirm that the structure being fit in the cable measurements by the extra terms is largely systematics, by noting that an antenna simulator (fifth panel) is fit more poorly for higher $w_{\rm terms}$, peaking at $w_{\rm terms}=5$. Regardless, for all these models, the inferred cosmic signal is reasonably consistent.} 
    \label{fig:wterms_resids}
\end{figure}

The first question is whether the iterative solution really does avoid being biased by the cable systematic. The proper way to answer this would be to construct a model for the cable measurements that included a flexible $\Gamma_{\rm src}$ systematic component, then determine the model with the highest Bayesian evidence.
However, choosing a flexible form for the (complex-valued) $\Gamma_{\rm src}$ that is able to capture the systematic is a rather involved task, and we defer it to future work\footnote{We can report that a simple polynomial scaling and delay is not a good model.}
In the meantime, we can gain some confidence by noting Fig. \ref{fig:wterms_resids}, which shows that different numbers of $w_{\rm terms}$ yield largely consistent inferred cosmic signals (bottom panel).
Indeed, also shown in this figure is the residuals to an antenna simulator -- a known input source with $\Gamma_{\rm src}$ designed to approximate the antenna itself. 
While this source is not used to fit the calibration, it can be used to check the results. 
Fig. \ref{fig:wterms_resids} indicates that $w_{\rm terms}=5$ (orange) -- the choice used in \np\ -- minimizes the its residuals. 
Higher $w_{\rm terms}$ decrease the performance of the antenna simulator, indicating that they are fitting systematics in the cable measurements themselves.
This is not a perfect test.
It is possible that the $w_{\rm terms}=5$ fit is partially biased by cable systematics, or that the antenna simulator itself has independent systematics in its $\Gamma_{\rm src}$ measurement. 
However, without performing a full investigation into the source and nature of the cable systematic, we can be reasonably confident that $w_{\rm terms}=5$ is providing a good, stable calibration.

The next question is how to define a likelihood that is able to restrict the cable measurements' impact on the noise-wave terms. 
In principle, this is impossible with a self-consistent likelihood. 
In this paper, we take the following approach: we first perform an iterative fit, and then, given the measured temperature of the cable inputs, we `decalibrate' to determine $\widehat{q}_{\rm src}$. To this, we add simulated Gaussian noise with variance determined empirically from the measurements. We then substitute these simulated cable `measurements' for the observed data.
In this way, to within correlations between the scale and offset and noise-wave parameters, we are guaranteed to obtain point-estimates of the noise-waves consistent with the iterative approach, with a posterior distribution consistent with the observed noise\footnote{In this work we use one specific noise realization for this simulated cable data. Given that we are in a high SNR regime, we do not expect results to be sensitive to the realization itself.}.

Table \ref{tab:evidence-unbiased} shows the Bayesian evidence computed by sampling models and data computed in this way, where the initial iterative calibration (to set the simulated cable data) was performed with the default $c_{\rm terms}=6$ and $w_{\rm terms}=5$ as used in \np. 
The highest Bayesian evidence is obtained for the fiducial number of terms.
For $w_{\rm terms}$ this is merely taken to be a consistency check of the code, as the highest evidence must be obtained for the $w_{\rm terms}$ used in the simulation.
However, this reasoning does not apply as strongly to $c_{\rm terms}$, since this is predominantly set by the \texttt{ambient} and \texttt{hot\_load} sources, which are not simulated.
Thus, the fact that we obtain the highest evidence for $c_{\rm terms} = 6$ is an indication that we truly require 6 terms for $T'_{\rm L}$ and $T'_{\rm NS}$\footnote{We note that using 6 terms also provides the best RMS on the antenna simulator, which is why it was chosen to be used in \np.}
This is further emphasised by the fourth column of Table \ref{tab:evidence-unbiased}, which shows the Bayesian evidence for models fit to data in which the simulated cable measurements were constructed based on fits with $(c, w) = (8, 5)$. 
Even in this case, the strongest evidence is obtained for a model with $c=6$, which is  a strong justification for our choice to use this number of terms for the rest of this paper.

\begin{figure}
    \centering
    \includegraphics[width=\linewidth]{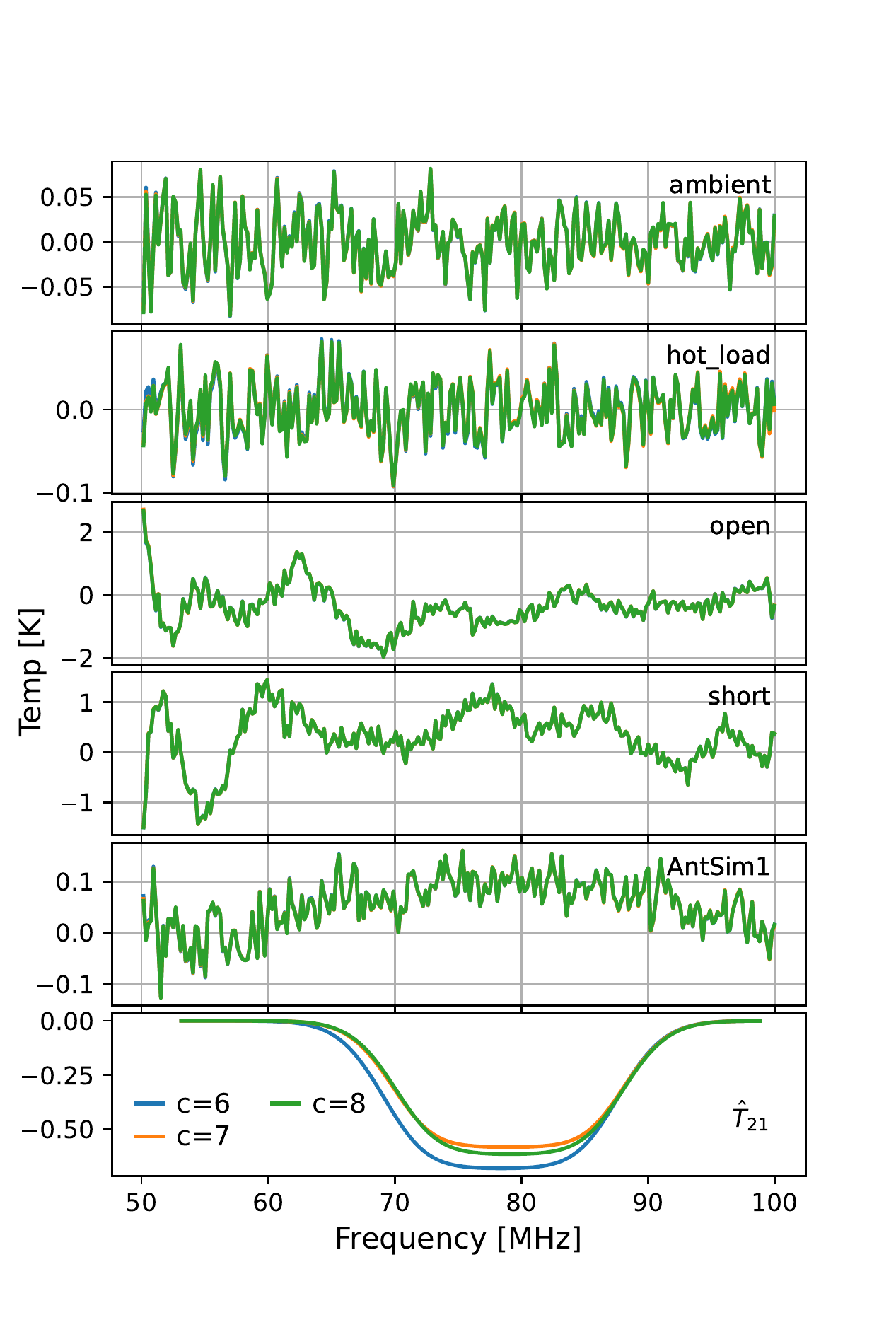}
    \caption{Calibration residuals for differing number of $c_{\rm terms}$, using the likelihood and data outline in \S\ref{sec:results-pure-calibration} -- i.e. in which simulated cable measurements are injected. The models shown correspond to the three with highest evidence in Table \ref{tab:evidence-unbiased}. All choices of $c_{\rm terms}$ produce similar estimates of the cosmic signal. 
    }
    \label{fig:resids_cterms}
\end{figure}

As a further check, we show the calibration residuals of the three highest-evidence models from Table \ref{tab:evidence-unbiased} in Fig. \ref{fig:resids_cterms}. 
As expected, each can perfectly reproduce the cable measurements, while differences in the loads and antena simulator are very small. 
There are noticeable differences in the inferred cosmic signal, however even these are stable to within the expected posterior reported in \np.

\begin{table}
    \centering
    \begin{tabular}{ccrr}
       \hline
         $\mathbf{c}_{\rm terms}$ & $\mathbf{w}_{\rm terms}$ & $\ln\mathcal{Z}$ [6,5] & $\ln\mathcal{Z}$ [8,5]\\
         \hline
             6  & 4 & -10065.5 &\\ 
         \textbf{6}  & \textbf{5} & \textbf{3083.7}   & \\
         6  & 6 & 3075.0    &\\
         \hline
         3 & 5 & -1787.6 &\\
         4 & 5 & 2944.7 &\\
         5 & 5 & 3029.1 & 3014.7\\
         \textbf{6}  & \textbf{5} &  \textbf{3083.7} & \textbf{3078.7} \\
         7  & 5 &  3075.4 & 3072.8\\
         8  & 5 &  3068.2 & 3068.3\\
         9  & 5 &  3062.2 & \\
         \hline
    \end{tabular}
    \caption{Bayesian evidence for pure calibration models (i.e. without field data) where the cable measurements are replaced by simulated data constructed from less-biased iterative solutions. The iterative solution uses $c=6$ and $w=5$. Since only the cable measurements are substituted for simulations, only the noise-wave parameters (set by $w_{\rm terms}$) are significantly affected by this choice. Thus the top portion of the table is a check that indeed the simulations yield a maximum evidence at the notional number of terms. The lower section modifies $c_{\rm terms}$ and thus provides justification for choosing $c_{\rm terms}=6$ to describe the calibration. We find that we obtain the maximum evidence for $c_{\rm terms}=6$ whether we use simulated cable data from a fit with $c'_{\rm terms}=6$ (third column) or 8 (fourth column).}
    \label{tab:evidence-unbiased}
\end{table}

In summary, while investigation into the source of the systematics in the cable measurements is a high priority for future work, the results of this section indicate that using an iterative approach to solve for the noise-wave parameters and $T'_{\rm L}$ and $T'_{\rm NS}$ produces results that maintain consistency between their respective inferred cosmic signals. 
We thus adopt this approach, wherein we use the iterative solutions to produce simulated cable data for our likelihood, for the remainder of this paper. Hereafter, we use this model with $c_{\rm terms}=6$ and $w_{\rm terms}=5$, which maximizes the Bayesian evidence.

\begin{figure}
    \centering
    \includegraphics[width=\linewidth]{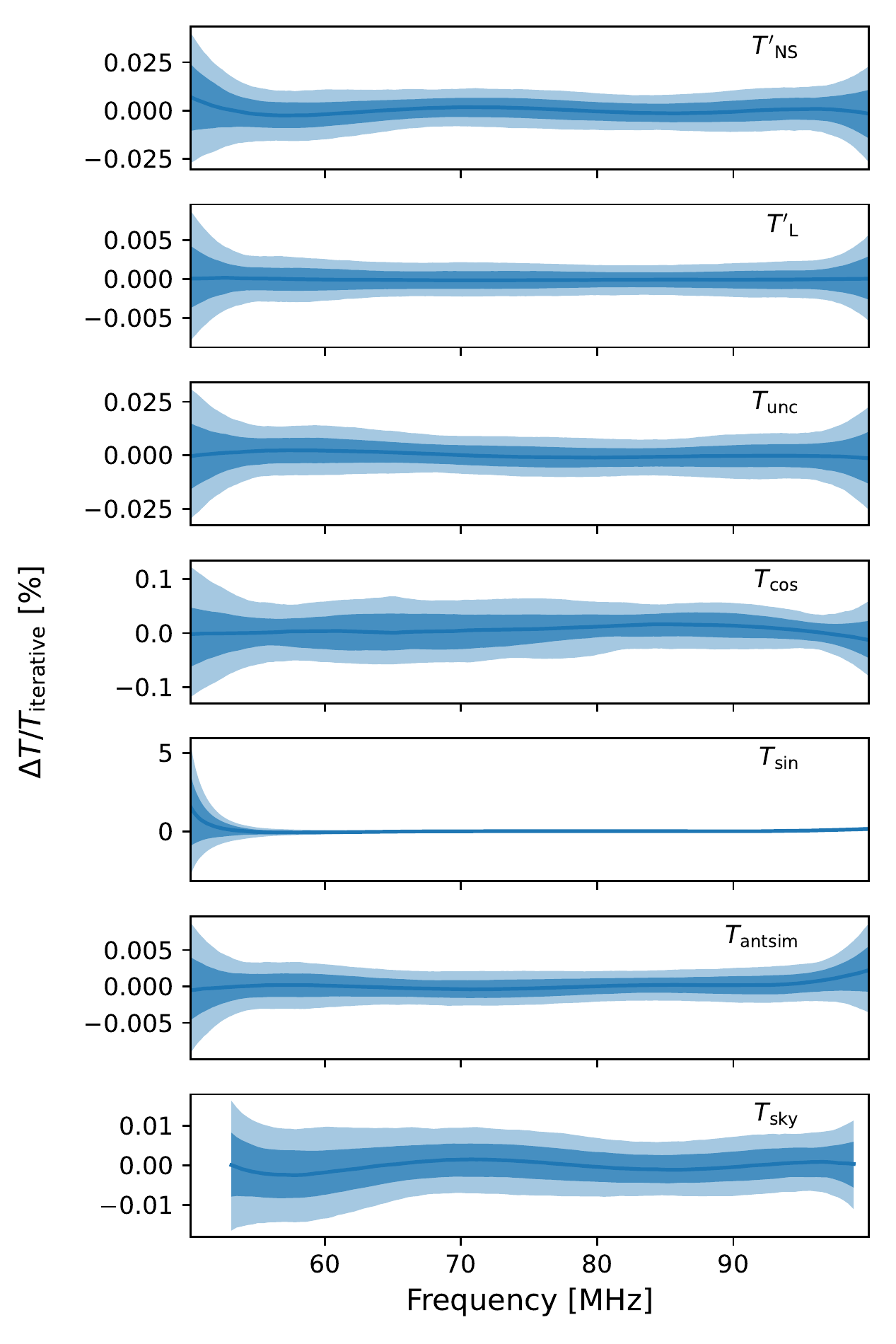}
    \caption{Posterior 1- and 2-$\sigma$ regions for the calibration parameters using the pure calibration likelihood described in \S\ref{sec:results-pure-calibration}. Each is shown as a fractional difference with respect to the iterative solution. Also shown is the posterior re-calibrated sky temperature, whose 1-$\sigma$ posterior fully agrees with the iterative solution, and is a hundredth of a percent in width.}
    \label{fig:calibration_regions}
\end{figure}

\begin{figure}
    \centering
    \includegraphics[width=\linewidth]{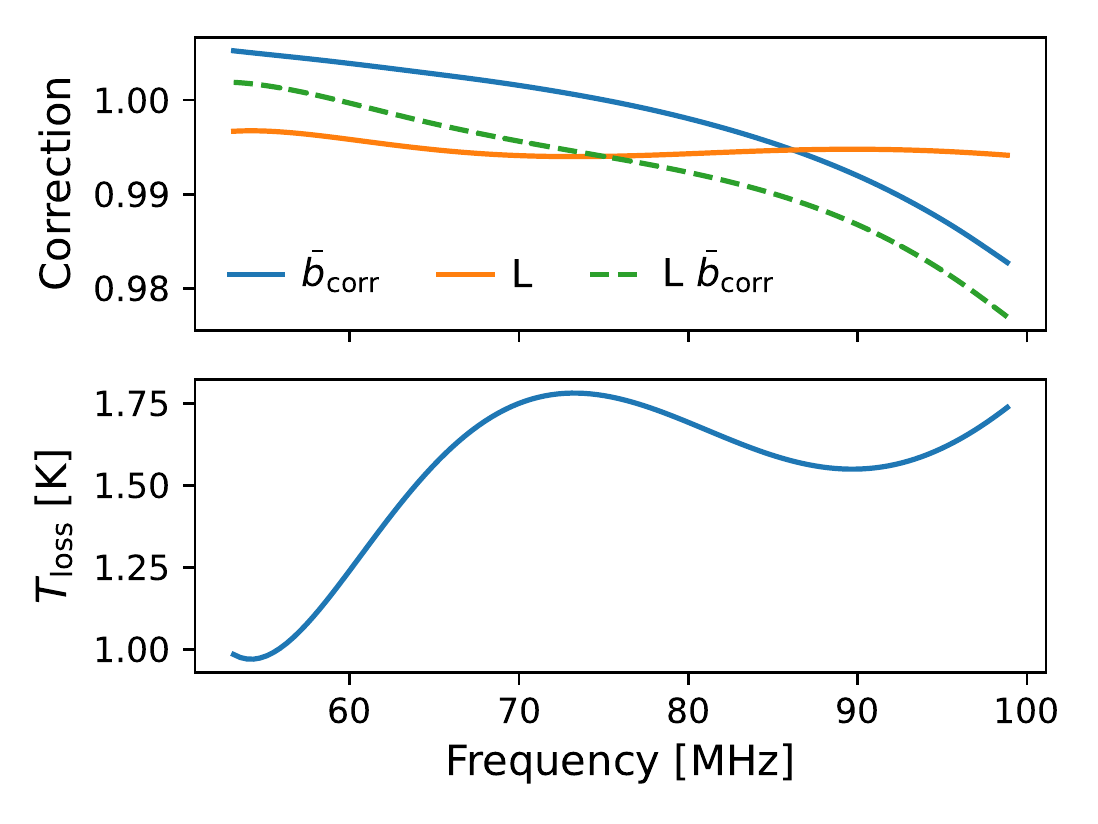}
    \caption{Various correction factors applied to the field data, as expressed in Eq. \ref{eq:processed_q}. In particular, the total loss (from antenna and ground) as well as the `beam correction' (Eq. \ref{eq:beamcorr}).}
    \label{fig:losses}
\end{figure}

Finally, we show the posterior distributions of the calibration parameters and calibrated antenna simulator and field data in Fig. \ref{fig:calibration_regions}.
We note that all calibration parameters have posteriors with width $\ll 1\%$, and are consistent with the iterative solutions, with the slight exception of $T_{\rm sin}$, which has a 5\% width at frequencies <53\,MHz.
The induced extra uncertainty on the sky data is of order 0.01\%. 
All curves exhibit higher uncertainty at the band edges, which is expected for the flexible polynomials we fit.

\section{A Probabilistic Sky Data Model}
\label{sec:field-data-model}
We now turn to derive a probabilistic model for data measured with the EDGES antenna in the field, which will include the previously-derived calibration likelihood as a subset.

As in the previous section, throughout this section, almost all of the quantities are frequency-dependent, and thus represented by a length-$N_\nu$ vector. 
However, since frequencies are not coupled in any of the modelling steps, we will omit their frequency-dependence in this section, and write each as a scalar (to be interpreted as a single element of a length-$N_\nu$ vector\footnote{This is in contrast to a continuous function of frequency, as it encodes not only the frequency dependence but also information about the frequency bin width.}).
We will explicitly note the rare quantities that are (assumed to be) independent of frequency.

The true average sky temperature is assumed to be (in absence of ionospheric distortions) merely a sum of foreground and cosmic signal:
\begin{equation}
    T_{\rm sky}(t) = T_{21} + \int \frac{{\rm d}\Omega}{2\pi}\ T_{\rm FG}(t, \Omega),
    \label{eq:sky_temp}
\end{equation}
where $t$ is the time of observation, and $\Omega$ is solid angle in the reference frame of the antenna (eg. azimuth and altitude) with the integral extending over the upper hemisphere.
We have made the assumption that the cosmic signal is isotropic and has negligible fluctuations on the scale of the horizon. 
This equation adopts the formalism in which the sky drifts across the frame of reference (i.e. $T_{\rm FG}$ changes with time with respect to $\Omega$).  

The EDGES antenna does not perform an unweighted integral over the sky; it is more sensitive to regions close to zenith, and this sensitivity pattern is defined by its primary beam, $B$.
Thus, EDGES in principle measures
\begin{equation}
    T_{\rm sky, beam}(t) = \int \frac{{\rm d}\Omega}{2\pi}\ B(\Omega) \left[T_{21} +  T_{\rm FG}(t, \Omega)\right].
    \label{eq:sky_temp_with_beam}
\end{equation}
The beam is normalized such that its integral is $2\pi$ over the upper hemisphere.
Notice that beyond an overall modification to the amplitude, the beam introduces distortions to the \textit{spectrum} of $T_{\rm sky, beam}$ compared to $T_{\rm sky}$.
This is true even if the beam is achromatic itself, as it couples spatial structure in the sky into spectral structure. This effect is commonly known as beam chromaticity.

Nevertheless, while detailed modelling of the beam-weighted foregrounds is an important line of inquiry \citep{Tauscher2020,Tauscher2020a,Mahesh2021}, when averaged over a wide range of LSTs, the beam chromaticity tends to average out, as demonstrated in \np\ by verifying consistency of cosmic signal estimate with and without beam correction.
Thus, in this paper, given that the FG temperature is an \textit{a priori} unknown smooth function of time and frequency, we simply replace the entire beam-weighted foreground term with a similar smooth function---allowing the higher-order terms of the unknown FG model to absorb any remaining structure from the beam---and correct for the beam chromaticity explicitly:
\begin{equation}
    T_{\rm sky,bc}(t)  \equiv  \frac{T_{\rm sky, beam}(t)}{b_{\rm corr}(t)} \approx  T_{21} + T_{\rm BWFG}(t),
    \label{eq:sky_temp_bcorr}
\end{equation}
where the `beam chromaticity correction' is given by \citep{Mozdzen2019}:
\begin{equation}
    b_{\rm corr}(\nu) = \frac{\int d\Omega \ B(\Omega, \nu) T_{\rm haslam}(\Omega, \nu_{\rm ref})}{\int d\Omega \ B(\Omega, \nu_{\rm ref}) T_{\rm haslam}(\Omega, \nu_{\rm ref})},
    \label{eq:beamcorr}
\end{equation}
with $\nu_{\rm ref} = 75\,{\rm MHz}$.
Note that this is an approximation; it accounts for the first-order frequency-dependent effects of the beam under the assumption that the cosmic signal is isotropic on the angular scales over which the beam modulates. For an achromatic beam and accurate sky model, this correction accounts for all chromatic structure leaked from angular scales to frequency. For realistic chromatic beams, there is unavoidably residual chromatic structure (after beam correction). This is why we denote the foreground term as ``$T_{\rm BWFG}$'', which indicates that the foregrounds we finally estimate must themselves account for this structure.

Throughout this work we use a phenomenological model for the 21\,cm signal during Cosmic Dawn as used in \np:
\begin{equation}
    T_{{\rm 21}, i} = - A\left(\frac{1 - e^{-\tau e^\psi}}{1 - e^{-\tau}}\right),
    \label{eq:flattened-gaussian}
\end{equation}
where
\begin{equation}
    \psi = \frac{4(\nu - \nu_0)^2}{w^2} \log \left[-\frac{1}{\tau} \log\left(\frac{1 + e^{-\tau}}{2}\right)\right]
\end{equation}
and the 21\,cm parameters are $\vth_{\rm 21} = [ A, \nu_0, w, \tau]$.


The antenna imposes additional frequency- and time-dependent gains on the incoming signal after the beam-convolution:
\begin{equation}
    T_{\rm sky, loss}(t) = L T_{\rm sky, beam}(t) + (1 - L)T_{\rm amb}(t),
    \label{eq:tsky_loss}
\end{equation}
where $T_{\rm amb}$ quantifies the \textit{frequency-independent} ambient temperature at the antenna-site at any time, and $L$ is considered here to be time-independent and encodes the product of losses incurred by the antenna, balun, connectors and ground:
\begin{equation}
    L = L_{\rm ant} L_{\rm balun}  L_{\rm conn} L_{\rm ground}.
    \label{eq:loss}
\end{equation}
In this work, we consider uncertainties in the modelling of the loss to be negligible. Its value is very close to unity for all frequencies.

Finally, the signal passes through the receiver, at which point a multiplicative gain and additive noise are imposed (cf. \S\ref{sec:calibration-model}), and we recognize that the entire signal chain (including the sky) has been stochastic:
\begin{equation}
    p_{\rm sky, meas}(t) = g(t) T_{\rm sky, loss}(t) + T_{\rm inst}(t) + n_{\rm sky, meas},
    \label{eq:fully_uncalibrated_temp}
\end{equation}
where $n_{\rm sky, meas}$ is a zero-mean Gaussian random variable. 

Applying the Dicke-switching and noise-wave formalism presented in \S\ref{sec:calibration-model}, we find the final measured three-position-switch power ratio is given by
\begin{equation}
    q_{\rm ant}(t) = \frac{T_{\rm sky,loss}(t) - T_{1}^{\rm ant}}{T_0^{\rm ant}} + n_{\rm ant}(t),
\end{equation}
with $T_0^{\rm ant}, T_1^{\rm ant}$ given by Eq. \ref{eq:t0t1}, and where we assume that $n_{\rm ant}$ is drawn from a zero-mean Gaussian distribution with variance $\sigma^2_{\rm q,ant}$(t)\footnote{Note that $n_{\rm ant}$ is different than $n_{\rm sky,meas}$ and while the latter is normally distributed to a very good approximation, the former is only Gaussian under the approximations outlined in \S\ref{sec:model:threepos}.}.

\subsection{Data Processing}
In \np\ the spectra from all times are averaged together. 
This clearly \textit{loses} information, and makes it more difficult to verify the truly global nature of the cosmological background \citep{Tauscher2020, Liu2014}, however without an accurate model of the low-frequency sky, it is necessary in order to average down systematics that decorrelate as the sky evolves.

Data-flagging and averaging was applied to an estimate of the \textit{sky temperature} rather than the raw measured quotients. That is, the data
\begin{equation}
    \widehat{T}_{\rm sky,bc}(t) = \frac{1}{L b_{\rm corr}(t)}\left[ \widehat{T}^{\rm ant}_0 q_{\rm ant}(t) + \widehat{T}^{\rm ant}_1 - (1 - L) T_{\rm amb}(t) \right]
    \label{eq:get_qant_fromT}
\end{equation}
are used to evaluate flags and perform averaging. Here, the estimated calibration functions were computed based on the iterative scheme outlined in \S\ref{sec:differences-iterative}, with $(c_{\rm terms}, w_{\rm terms})=(6,5)$.

The final averaged spectrum is
\begin{align}
    \widehat{\overline{T}}_{\rm sky,bc} &= \frac{1}{w} \sum_j \xi(t_j) \widehat{T}_{\rm sky,bc}(t_j), \\
    &\approx \frac{1}{L \bar{b}_{\rm corr}} \left[ \widehat{T}^{\rm ant}_0  \overline{q}_{\rm ant} + \widehat{T}^{\rm ant}_1 - \overline{T}_{\rm loss} \right],
    \label{eq:processed_q}
\end{align}
where $\overline{T}_{\rm loss} \equiv (1 - L)\overline{T}_{\rm amb}$ is the mean loss temperature, $\xi \in \{0,1\}$ are the per-frequency flags at each time-stamp, and $w = \sum_j \xi(t_j)$ is the number of (unflagged) samples per-frequency (over all measured time samples), and the effective quotient, ambient temperature and beam correction are given by weighted average over time samples,
\begin{equation}
    \overline{X} = \frac{1}{w} \sum_{j} {\xi}(t_j) X(t_j).
\end{equation}
Note the the second relation (Eq. \ref{eq:processed_q}) is an approximation, due to the fact that the beam correction term is time-dependent and the mean of a product is the not the product of means. Nevertheless, we expect this effect to be small, and delay its proper treatment to future work.

Note that $\widehat{\overline{T}}_{\rm sky,bc}$ is precisely the publicly-available spectrum shown in Fig. 1 of \np.

Our likelihood does not use $\widehat{\overline{T}}_{\rm sky,bc}$ directly, but instead uses the more basic quantity $\overline{q}_{\rm ant}$. This has the benefit of being only slightly dependent on the \textit{estimates} $\widehat{T}^{\rm ant}_0, \widehat{T}^{\rm ant}_1$. This slight dependence arises through the fact that the flags, $\vect{\xi}$, are computed based on the calibrated data. This dependence is extremely insensitive to small changes in the calibration temperatures, because the flags are empirically assigned based on a non-parametric estimate of whether the datum is an outlier (over either the frequency or LST axis). Changes in the calibration temperatures introduce very smooth changes in the data as a function of frequency/LST, and therefore are highly unlikely to change the flags. 
We obtain $\overline{q}_{\rm ant}$ simply by inverting Eq. \ref{eq:processed_q} using the publicly-available data $\widehat{\overline{T}}_{\rm sky,bc}$ and values for $L$, $\bar{b}_{\rm corr}$, $\overline{T}_{\rm loss}$ and $\hat{T}^{\rm ant}_{0,1}$ obtained directly from the \np\ analysis code.

\subsection{Data Model}
The antenna-sourced power quotient, $q_{\rm ant}$ can be treated on the same footing as the input calibration sources in the calibration likelihood, Eq. \ref{eq:cal_likelihood_conceptual}, i.e. by expanding the set of sources summed over to $\mathcal{S}_{\rm joint} = \mathcal{S}_{\rm cal} \cup \{{\rm ant} \}$.
Just like the other input sources, its distribution is assumed to be well-approximated by an uncorrelated multivariate Gaussian, with mean $\langle \bar{q}_{\rm ant} \rangle$ and variance $\sigma^2_{\rm q_{\rm ant}}$.

In contrast to the other sources, however, we do not have a low-noise measurement of the true input temperature of the antenna, $T_{\rm src} \equiv T_{\rm ant}$.
Instead, we have a \textit{model} for the true temperature:
\begin{align}
    T_{\rm ant} &=  L \bar{b}_{\rm corr} \left[\overline{T}_{\rm BWFG} + T_{21}\right] +
    \overline{T}_{\rm loss} \\
    &\equiv T^{\rm ant}_0 \langle \overline{q}_{\rm ant} \rangle +  T^{\rm ant}_1.
    \label{eq:expectation_qant}
\end{align}
where $\overline{T}_{\rm BWFG}$ models the time-averaged beam-weighted foregrounds, and should be a smooth function of frequency.


The variance is in principle an unknown function that should be modelled and inferred. 
However, in this work we simply estimate the variance by analysis of the residuals of the data to high-order smooth models.
Note that this variance model is different to that use in \np, who assumed a frequency-independent variance\footnote{The magnitude of this variance in \np\ was not important for the main results as they were merely maximum likelihood estimates.}.

We assume a very spectrally smooth, but otherwise flexible, model for the time-averaged beam-weighted foregrounds, which allows it to absorb potential calibration and beam chromaticity errors so long as they are spectrally distinct from the expected 21\,cm signal.
The model we employ here is colloquially termed the \textsc{linlog} model:
\begin{equation}
    \overline{\vect{T}}_{\rm BWFG} = \left(\frac{\vect{\nu}}{\nu_0}\right)^{-2.5} \sum_i^{N_{\rm FG}} \vth_{{\rm FG}, i} \ln\left(\frac{\vect{\nu}}{\nu_0}\right)^i = \matr{\Phi} \vth_{\rm FG},
    \label{eq:linlog}
\end{equation}
where $\matr{\Phi}$ is the $N_\nu \times N_{\rm FG}$ matrix of \textsc{linlog} basis functions.
Note that this is equivalent to assuming a \textsc{linlog} model of foregrounds of the same order at each time $t$, where $\vth_{{\rm FG},i} = \int \vth_{{\rm FG},i}(t) dt$. 

With these models, we have a joint calibration and sky model likelihood that is a simple extension of the pure-calibration likelihood (cf. Eq. \ref{eq:cal_likelihood_conceptual}):
\begin{equation}
    \mathcal{L}_{\rm joint}(\vth_{\rm mdl}, \vth_{\rm 21}, \vth_{\rm FG}) = \mathcal{L}_{\rm cal}(\vth_{\rm mdl}) \cdot \mathcal{L}_{\rm ant}(\vect{q}_{\rm ant} | \vth_{\rm mdl}, \vth_{\rm 21}, \vth_{\rm FG}),
    \label{eq:field_likelihood_conceptual}
\end{equation}
where $\mathcal{L}_{\rm ant}$ is exactly Eq. \ref{eq:src_likelihood_conceptual} with src=ant.

Similar to the pure-calibration likelihood, we can also represent the joint likelihood in a way that highlights the linear parameters, so that we can use the AMLP method described in \S\ref{sec:amlp}. We give this representation in App. \ref{app:linear-field-model}. Briefly, the \textsc{linlog} foreground parameters are linear (along with the $T_{\rm lin}$), while $\vth_{\rm NS}$ and $\vth_{21}$ are non-linear.

\section{Joint Calibration and Sky Model Results}
\label{sec:results}
\begin{figure}
    \centering
    \includegraphics[width=\linewidth]{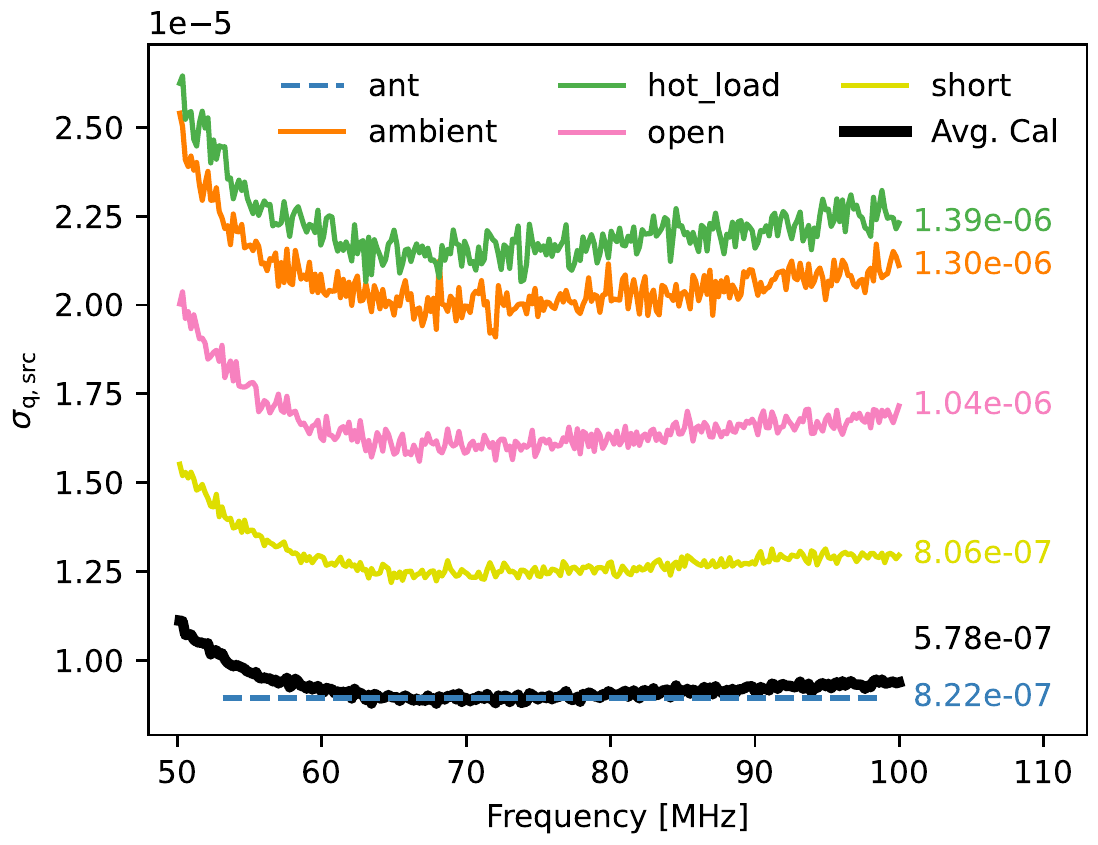}
    \caption{Empirical thermal uncertainty of the different input sources, $\sigma_{\rm q, src}$, including the antenna measurements from the field (blue). The thicker black line is the mean uncertainty over all calibration sources, i.e. $\sqrt{\sum_{\rm src} \sigma^2_{\rm q, src}}/N_{\rm src}$. Numbers to the right of each curve are the mean uncertainty over frequency for the given source. While the field antenna measurements individually have the lowest per-channel uncertainty (at the displayed frequency resolution), the combination of all calibration data has slightly lower total mean uncertainty (5.78$\times10^{-7}$ vs. $8.22\times10^{-7}$). Note that both calibration and field data have been averaged over a different amount of time and number of frequency channels before being displayed, making the numbers to the right the only `comparable' figure.}
    \label{fig:thermal_uncertainty}
\end{figure}

We first answer the question of the relative thermal uncertainty between the calibration data and field data. Fig \ref{fig:thermal_uncertainty} shows the thermal noise, $\sigma_{\rm q, src}$, on each input source as a function of frequency. While each individual calibration source has a higher uncertainty per frequency channel, the combination of higher frequency resolution and multiple calibration sources results in a slightly lower overall calibration uncertainty when averaged over all frequency channels and sources (black curve and number).
Thus, we naively expect the calibration data to be slightly more constraining than the field data.

Next, we turn to the posteriors of the joint calibration and sky data likelihood. 
Fig. \ref{fig:absorptions} shows the resulting Bayesian posteriors when the likelihood defined in Eq. \ref{eq:field_likelihood_conceptual} is applied to the full set of available data (i.e. both lab calibration data as discussed in \S\ref{sec:results-pure-calibration} and the averaged sky spectrum as given in Eq. \ref{eq:processed_q}). 
In fact, the figure shows these posteriors as the colored regions, while also showing posteriors from an `isolated' sky model fit as grey hashed regions. The `isolated' fit is obtained simply by using the sky data alone, where the data has been pre-calibrated using the \textit{maximum a posteriori} point from our calibration model.
Thus, this plot reveals the impact of performing a joint `calibration and sky model' fit on the cosmic inference, as compared to the traditional process of choosing a calibration and then performing the sky model fit in isolation.

We look for two things: bias and posterior spread differences.
Note that the regions shown are the 68\% and 95\% (i.e. 1- and 2-$\sigma$) quantiles, while the solid/dashed lines are the median value. 
Considering biases between the two approaches, we note that for low $N_{\rm FG} \le 6$, the posterior regions are discrepant to varying degrees: extremely so for $N_{\rm FG}=4$, with slightly better agreement ($\sim 1\sigma$) for $N=5$, and $2\sigma$ for $N_{\rm FG}=6$. 
Conversely, for $N_{\rm FG}>6$, we find extremely good agreement between the approaches, with almost perfect overlap of their distributions. 
The reason for this is likely that the sky-averaged data contains non-cosmic structure that is unable to be fit by a \textsc{linlog} model with fewer than 7 terms. For the isolated approach, this simply causes the cosmic inference to be biased, as it is correlated with the foreground model. 
For the joint model, the effect is partially ameliorated by the calibration itself absorbing some of the extra structure. 

This is made clear in Fig. \ref{fig:posterior_tns}, which shows the posteriors on the polynomial coefficients of the calibration scaling parameter, $T'_{\rm NS}$. In that figure, we show only $N_{\rm FG}=(4,6,8,10)$ for visual clarity. 
The grey-dashed cross-hairs show the results of the traditional iterative solution, on which the calibration likelihood is based. 
As speculated based on Fig. \ref{fig:absorptions}, the low-$N_{\rm FG}$ posteriors are discrepant with the iterative solutions, while the high-$N_{\rm FG}$ posteriors are consistent with them.
On its own, this merely indicates that the calibration solutions are biased when they need to fit out structure in the sky data that the sky model itself is too inflexible to deal with.
However, comparing this to Fig. \ref{fig:absorptions} reveals that the \textit{manner} in which the calibration solutions is biased is precisely to absorb the residual smooth FG structure, allowing the cosmic signal to remain at its preferred location consistent with \np. 
This is further clarified in Fig. \ref{fig:linear_posteriors}, which shows the posteriors for $T^{\rm ant}_0$ and $T^{\rm ant}_1$ as a comparison to the iterative solution, for $N_{\rm FG} \in (4, 10)$. While the additive temperature is almost unaffected by the change in foreground model complexity, the multiplicative temperature is altered in the $N_{\rm FG}=4$ case, with a noticeable dip of magnitude 0.02\% around 75\,MHz. 
This lends weight to the proposition that the cosmic feature is truly preferred by the data, as it appears stable with respect to $N_{\rm FG}$ and is strong enough to influence the calibration to remain stable.

\begin{figure}
    \centering
    \includegraphics[width=\linewidth]{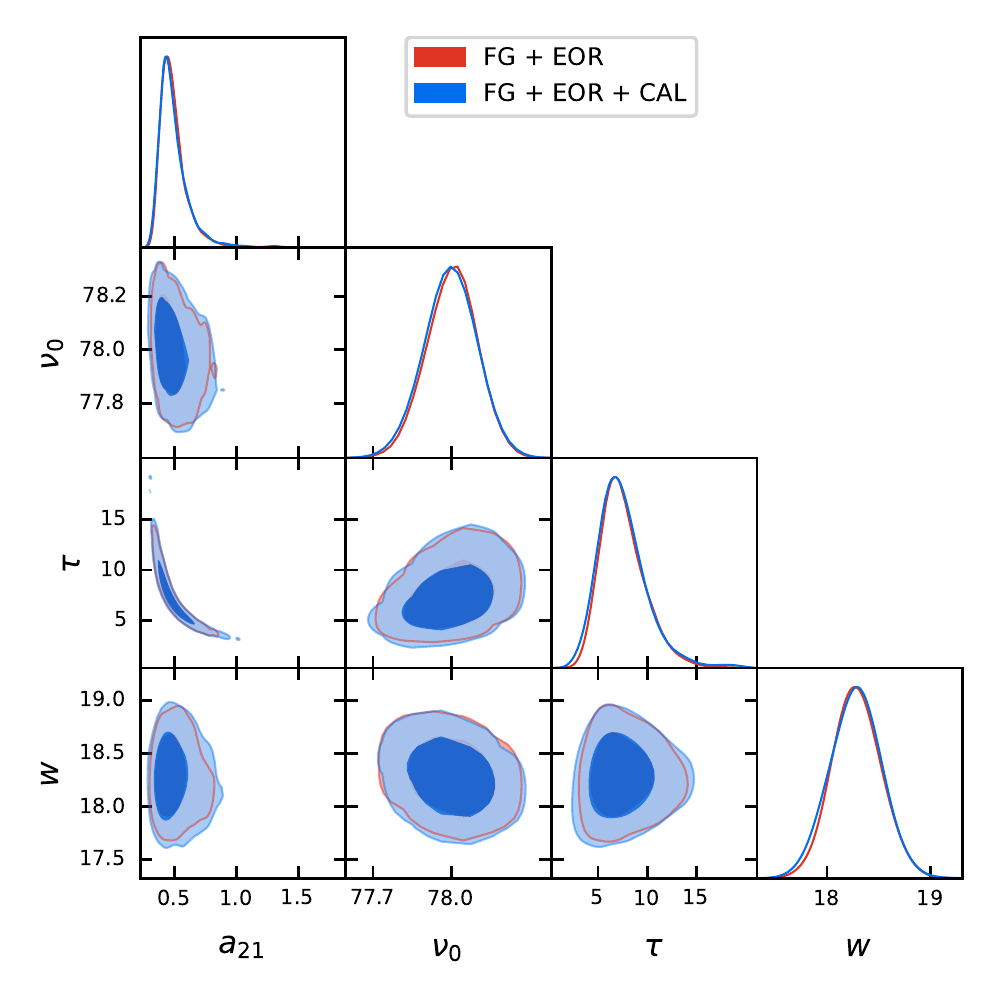}
    \caption{Posteriors on the 21\,cm flattened-gaussian absorption model, Eq. \ref{eq:flattened-gaussian}, from both isolated and joint fits for $N_{\rm FG}=10$. Here, the foreground model is complex enough to describe the non-cosmological components of the data, and the calibration model is not required to account for the beam-weighted foregrounds. This results in very similar predictions for the absorption parameters (cf. Fig. \ref{fig:absorptions}). We also see very similar posterior spread in the parameters, due to the fact that the calibration uncertainty is extremely low (cf. Fig. \ref{fig:calibration_regions}).}
    \label{fig:t21params}
\end{figure}

\begin{figure*}
    \centering
    \includegraphics[width=\linewidth]{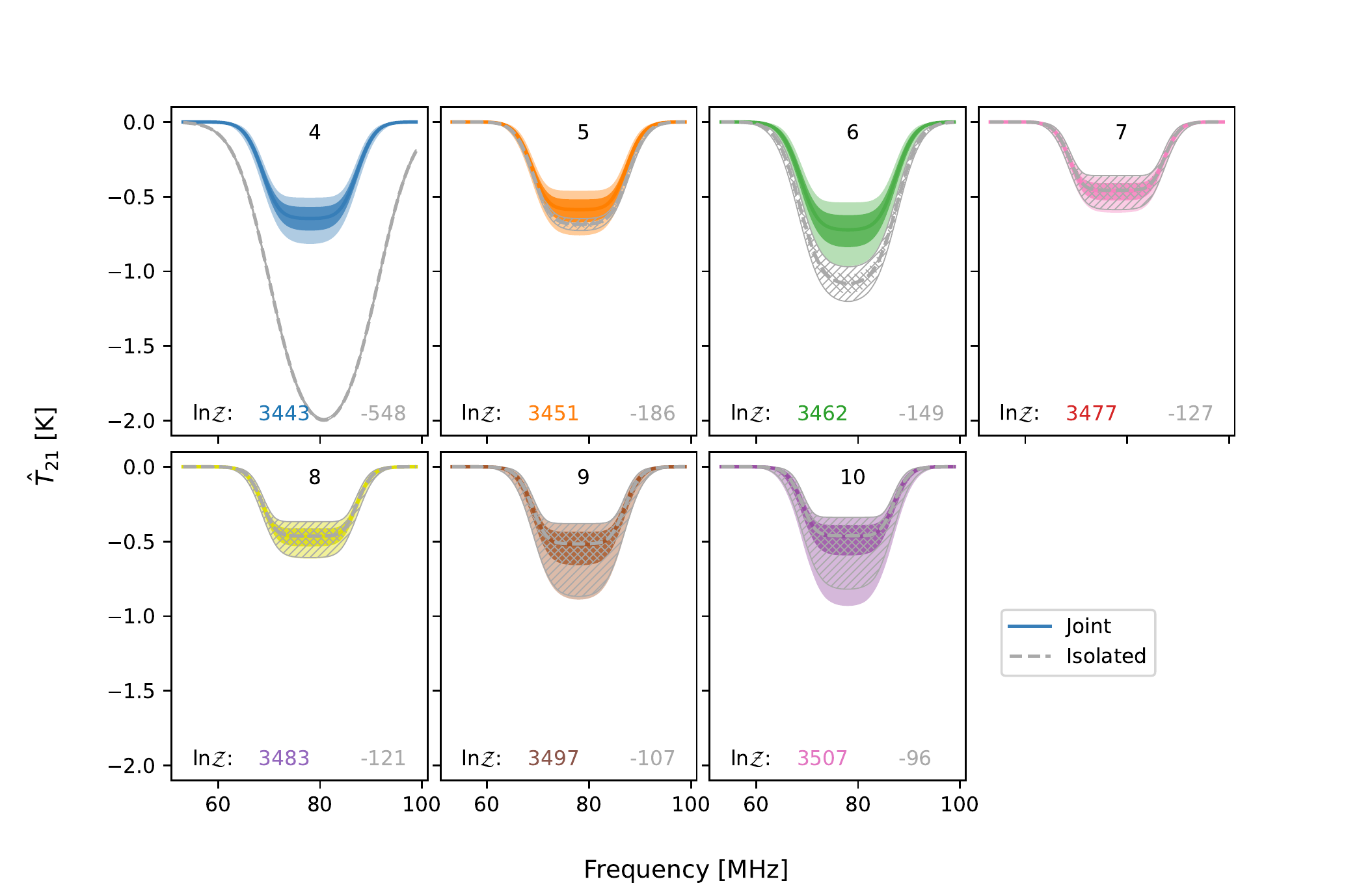}
    \caption{The inferred posteriors of the 21\,cm signal using different numbers of FG terms (number at top of each panel). Colored regions show 1- and 2-$\sigma$ quantiles from a joint calibration and sky model inference, while the grey hashes show the same for an `isolated' fit of the sky model to pre-calibrated sky data.
    For $N_{\rm FG} \le 6$ the joint and isolated fits are discrepant, due to the FG model being insufficiently flexible to fit the data. For $N_{\rm FG}\ge7$, the posteriors are remarkably similar, indicating that the calibration uncertainty is minimal.
    }
    \label{fig:absorptions}
\end{figure*}

\begin{figure*}
    \centering
    \includegraphics[width=\linewidth]{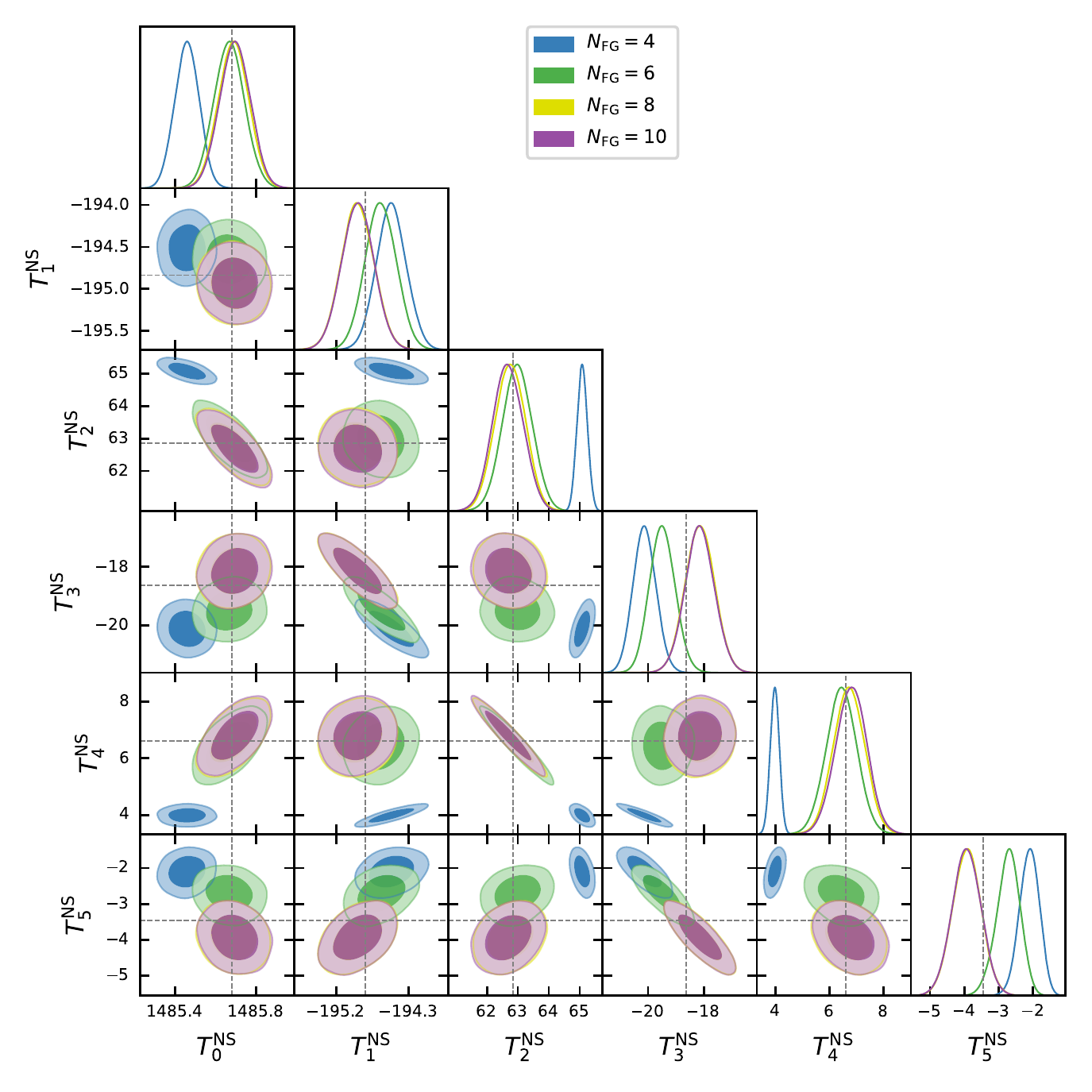}
    \caption{Posteriors for the calibration function $T'_{\rm NS}$ after running the joint calibration and sky model inference for different numbers of FG terms. Grey dashed cross-hairs are the iterative solutions for the fiducial calibration model. Using a flexible FG model lets the calibration remain close to the best fit from purely lab-calibration data. The calibration absorbs some of the sky data structure when the sky model itself is not flexible enough. See Fig. \ref{fig:linear_posteriors} for a projection of this posterior to frequency-space.}
    \label{fig:posterior_tns}
\end{figure*}

\begin{figure}
    \centering
    \includegraphics[width=\linewidth]{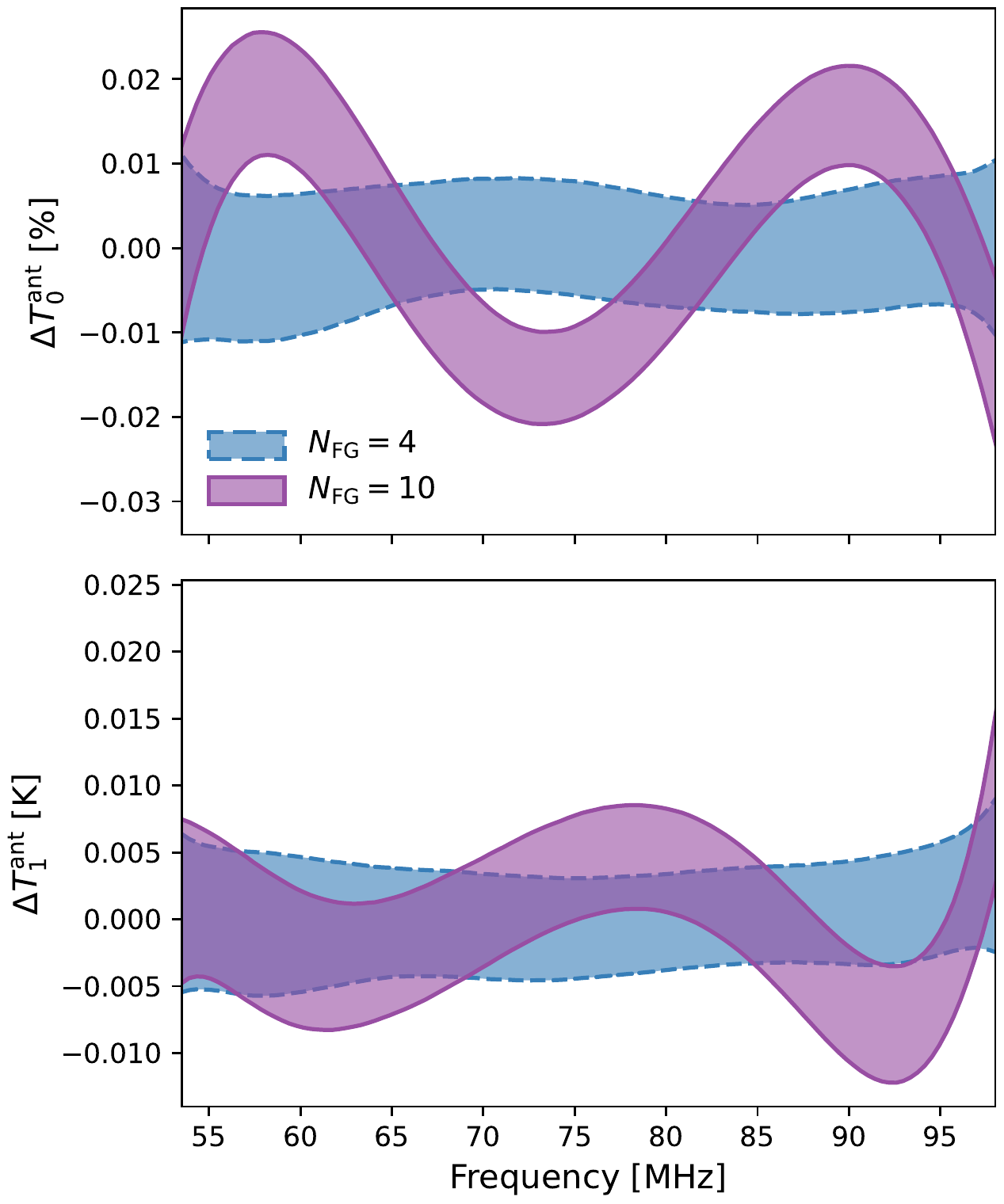}
    \caption{Posteriors of linear calibration parameters, $T^{\rm ant}_0$ and $T^{\rm ant}_1$ for the joint calibration and data likelihood. Shown are the posteriors for $N_{\rm FG}=4$ and $N_{\rm FG}=10$, which are represented by the same colors in Figs. \ref{fig:absorptions} and \ref{fig:posterior_tns}. The curves shown are in comparison to the iterative solution with calibration data only. The multiplicative gain, $T_0$ is shown as a fractional difference, while the additive temperature, $T_1$, is shown as an absolute difference. }
    \label{fig:linear_posteriors}
\end{figure}

In terms of the posterior spread, we note the general trend in both approaches for the spread to increase as $N_{\rm FG}$ increases. 
This is to be expected, since the extra flexibility leaves more room for the cosmic signal to vary. 
This trend is broken only by $N_{\rm FG}=6$, which is an outlier for both its median estimate and its posterior spread. 
It is unclear what the precise source of this anomaly is, though it is not of much importance, since  $N_{\rm FG}=6$ is not the best model (from the Bayesian evidence, see below).
Finally, we note that for $N_{\rm FG}>6$, the spread of both approaches is extremely similar.
This is to be expected, since Fig. \ref{fig:calibration_regions} shows such a small (0.01\%) spread in the calibrated sky data posterior from the calibration alone. 
That is, as long as the FG model is sufficiently flexible so that the calibration is not drawn away from the lab data, the uncertainty on the calibration itself does not add significant uncertainty to the cosmic inference.
This is reinforced by Fig. \ref{fig:t21params}, which shows the posteriors on $\vect{\theta}_{21}$ for both the isolated and joint likelihoods, for $N_{\rm FG}=10$.
The posteriors are almost identical, implying that the uncertainty on the calibration parameters is extremely small.

A final point to note is that Fig. \ref{fig:absorptions} also lists the Bayesian Evidence, ${\rm ln}\mathcal{Z}$, for each of the models, 
For both the joint and isolated fits, the evidence increases indefinitely with $N_{\rm FG}$.
This strongly indicates that there is structure in the sky data that the flattened-Gaussian signal and \textsc{linlog} FG model are insufficient to account for. 
While the evidence \textit{may} peak for even higher $N_{\rm FG}$, it is unclear that these would be preferred, given our prior of spectrally smooth foregrounds. 
It is much more likely that the extra residual structure is due to residual beam chromaticity or inaccuracies in the measurement of the antenna's reflection characteristics. 
Pursuing the constraint of such systematics is planned for future work.

\section{Conclusions}
\label{sec:conclusions}

In this paper, we have developed a Bayesian likelihood for the joint estimation of receiver calibration parameters, foregrounds and 21\,cm signal for the EDGES global experiment. 
Our approach is similar to that of \citep{Roque2020}, except that, with an eye towards including more sophisticated systematics in the future, we do not utilize conjugate priors, but instead improve efficiency by marginalizing over our many linear parameters \citep{Monsalve2018,Tauscher2020}.
We applied this joint fit to data from the first reported evidence of a detection of the global 21\,cm signal in \citet{Bowman2018}.

Our first investigation was for data taken purely in the lab, meant to inform the receiver calibration alone.
This data consists of spectra and reflection parameters from four sources with known temperature.
Two of the sources are designed to have very low reflections, while the other two -- the open and shorted cable -- have large reflections, designed to probe the noise-wave temperatures.
We found that a significant systematic exists in the two cable measurements, resulting in $\sim 20-30\sigma$ residuals in those measurements after calibration. 
Such systematics were entirely absent from the other two measurements, whose reflections were small (i.e $|\Gamma_{\rm src}| \ll 1$).
We found that the these systematics significantly bias resulting cosmic signal estimates when calibration is performed using a likelihood that treats all input sources on the same footing.
However, the established iterative solution technique \citep{Monsalve2017} is able to largely avoid this bias by restricting the influence of the cable measurements.
Despite the cable systematics, applying the iterative calibration solutions to an `antenna simulator' designed to mimic the reflection characteristics of the EDGES antenna yields reasonable residuals ($\sim2\sigma$), and the effect of added calibration flexibility on cosmic inference is minimal.
Thus, to avoid this bias in our Bayesian likelihood, we adopt an approximate method in which we use the iterative solutions to \textit{simulate} cable measurements without systematic biases, adding Gaussian noise consistent with empirical estimates. 
Using this method, we find that using 6 terms for the scale and offset temperatures (i.e $c_{\rm terms}=6$) maximizes the Bayesian evidence, irrespective of the number of $c_{\rm terms}$ used in the iterative solution. 
This confirms the choice of \np, where this number of terms was chosen based on residuals of the antenna simulator. 

We then performed a joint fit of our calibration model along with a sky model consisting of  \textsc{linlog} foregrounds and a flattened-Gaussian cosmic signal. 
We were careful to include all other losses and corrections applied in \np, including beam correction, ground loss and balun loss. 
We found that our joint model infers a cosmic signal consistent with \np, for $N_{\rm FG} = 4-10$. 
This is in contrast to an `isolated' inference in which the `best-fit' calibration is applied to the sky data and the sky model is fit alone -- in this approach only high-$N_{\rm FG}$ fits are consistent in their predictions.
This, along with the rising Bayesian evidence with $N_{\rm FG}$, indicates that there is structure in the sky that requires $N_{\rm FG}>6$ to begin to capture. Nevertheless, the inferred feature is strong enough that in the joint fit, the calibration tends to absorb the extra structure that the foregrounds are unable to fit, keeping the cosmic inference consistent between different numbers of foreground terms.

One question that naturally arises in the context of this work is whether a full joint model is necessary. 
We have shown that, under the calibration assumptions made in this work, a joint model is unnecessary \textit{if the foreground model is sufficiently complex to describe the sky data}. 
The uncertainty on the calibration parameters is small enough that the extra uncertainty propagated to the cosmic signal parameters is negligible (cf. Fig \ref{fig:thermal_uncertainty}).
However, if the foreground model is too inflexible to account for the sky data, a joint model is more robust. 
Nevertheless, it would seem more appropriate to set the foreground model to be sufficiently flexible, and use an isolated fit, rather than a joint fit. 
This may not remain true as further calibration uncertainties are included. 

A further question is whether the joint model presented here may lend itself to more bias than an isolated model. 
Such a conclusion suggests itself as a possibility upon consideration of the pure calibration likelihood developed in this work. 
In that likelihood, we necessarily combined data from all calibration sources, weighted according to their thermal uncertainty. However, since our model for \textit{some} of those sources was incomplete (i.e. the cable measurements), this leaked bias into the models for all sources. 
In this case, `isolating' the measurements and models reduces the overall bias.
This reasoning may also be the case for the sky model and data.
By letting the calibration models ``see'' the sky data in the joint model, they are able to be influenced by it. This is not a problem, and indeed is the correct thing to do, if our sky model is accurate. However, if it is not, the calibration solutions will be pushed away from the solutions they would obtain purely from calibration data. Whether this is really a problem hinges on one's prior credence on the sky model's accuracy. If one is very confident in the sky model, then it is perfectly appropriate for the calibration model to be moved away from the lab data by the sky. If not, then it is appropriate to conclude that such movement is systematic bias. 
In this work we established that while the foreground and calibration models are correlated, the sharp features of our deep flattened-gaussian model are sufficiently uncorrelated with either such that its estimate remains constant against their changing complexity. 
In principle, model selection will play the crucial role of deciding which model is most appropriate.

We note that while the full joint model presented here is potentially unnecessary, in the sense that the posteriors for the parameters of interest are not significantly affected by including the calibration parameters, the Bayesian formalism for the calibration itself has proven to be highly useful.
The bias in the calibration solutions for the cable measurements, noted and discussed in \S\ref{sec:differences-iterative}, is only able to be properly diagnosed under the Bayesian framework presented here. Comparing Bayesian Evidence between different systematics models should be an effective solution to understanding where the bias comes from -- a problem we leave to future work.

\subsection{Future Work}
The results in this paper represent the foundation of a broader program that will be required to verify the results of \np. 
The foundation is the Bayesian statistical framework, in which various systematic biases and uncertainties can be added and jointly inferred along with the cosmic signal. 
In this paper, we have merely focused on the `easiest' of these uncertainties -- the receiver calibration -- as an initial exploration.
Five additional instrumental systematics are candidates for future modelling:
\begin{enumerate}
    \item The calibration cable measurements $\Gamma_{\rm short/open}$, which are behind the bias seen in Fig. \ref{fig:lab-resids}. 
    \item $\Gamma_{\rm inst}$ and $\Gamma_{\rm ant}$, for which we have multiple measurements taken over multiple years (in-situ for $\Gamma_{\rm ant}$).
    \item The beam chromaticity, for which the beam model itself could be better characterized, as well as the formalism in which the correction is applied.
    \item The various losses involved: antenna, ground, balun etc.
\end{enumerate}
In particular, the cable measurement bias will be an important systematic to characterize, in order to enable a more self-consistent likelihood.


\section*{Acknowledgements}
This work was supported by the NSF through research awards for EDGES (AST-1609450, AST-1813850, and AST-1908933). N.M. was supported by the Future Investigators in NASA Earth and Space Science and Technology (FINESST) cooperative agreement 80NSSC19K1413. PHS was supported in part by a McGill Space Institute fellowship and funding from the Canada 150 Research Chairs Program. EDGES is located at the Murchison Radio-astronomy Observatory. We acknowledge the Wajarri Yamatji people as the traditional owners of the Observatory site. We thank CSIRO for providing site infrastructure and support.  \textit{Software}: This paper has made use of many excellent software packages, including \textsc{numpy} \citep{Harris2020}, \textsc{scipy} \citep{Virtanen2020}, \textsc{polychord} \citep{Handley2015,Handley2015a}, \textsc{astropy} \citep{Robitaille2013,AstropyCollaboration2018}, \textsc{getdist} \citep{Lewis2019}, \textsc{matplotlib} \citep{Hunter2007}, \textsc{h5py} \citep{Collette2013} and \textsc{yabf}\footnote{\url{https://github.com/steven-murray/yabf}}. 

\section*{Data Availability}

All publicly-available data used in this work, as well as analysis software in the form of Jupyter notebooks, can be accessed at \url{https://github.com/edges-collab/bayesian-calibration-paper-code}. Raw calibration data is available on reasonable request via email to the corresponding author. Software used to perform work with this data is publicly available at \url{https://github.com/edges-collab}. 
Output data products, such as MCMC chains, resulting from this work are also available upon reasonable request to the corresponding author.



\bibliographystyle{mnras}
\bibliography{library} 


\appendix

\section{Expectation of a Ratio of Gaussian Variables}
\label{app:stat-identities}

For random variables $X, Y$, where the density of $Y$ at zero is negligible, the expectation of the ratio can be approximated via Taylor series to second-order as
\begin{align}
    \label{eq:stat:expectation-ratio}
    E[X/Y] \approx \frac{E[X]}{E[Y]} \left(1 - \frac{{\rm Cov}(X, Y)}{E[X]E[Y]} + \frac{{\rm Var}(Y)}{E[Y]^2}\right).
\end{align}

\section{Analytic Marginalization of Linear Parameters}
\label{app:marginalized-likelihood}
MCMC techniques are typically inefficient when dealing with large numbers of parameters, due to the curse of dimensionality. 
In this appendix, we review the derivation of a technique used to reduce the effective dimensionality of the model to be explored, via analytical marginalization over some of the parameters.
We note that this technique is not new, even in the context of global experiments \citep[eg.][]{Lentati2017,Monsalve2018,Tauscher2021}. 
However, at first glance, these papers suggest differing results (i.e. they conclude with different formulae that are not obviously identical). 
In this appendix, we derive the same results, and show that they are not in disagreement.

The technique is applicable when the following conditions hold (cf. \citet{Tauscher2021}):
\begin{enumerate}
    \item The likelihood of the data, $\vect{d}$, is Gaussian, i.e. 
    \begin{equation}
        \mathcal{L}(\vect{d} |\vth) \propto \left| \matr{\Sigma}^{-1}(\vth)\right| \exp\left\{-\frac{1}{2} \vect{r}^T \matr{\Sigma}^{-1}(\vth) \vect{r} \right\},
    \end{equation}
    where the $\vect{r} = \left(\vect{d} - \vect{m}(\vth)\right)$ is the residual of the data to a model, $\vect{m}$, dependent on the parameters $\vth$ and evaluated at the same coordinates as the data, and $\matr{\Sigma}$ is a model for the covariance of the data, potentially dependent on the parameters as well. 
    \item After specification of the values of a \textit{subset} of the parameters, to be called the non-linear parameters $\vth_{\rm NL}$, the model is linear in the remaining parameters, $\vth_{\rm L}$. That is, the parameters can be split into two groups, $\vth = \{\vth_{\rm L}, \vth_{\rm NL}\}$ such that when the model is conditioned on $\vth_{\rm NL}$, the gradient of the model with $\vth_{\rm L}$ is independent of $\vth_{\rm L}$.
    \item The covariance depends only on the non-linear parameters.
    \item The priors on the two sets of parameters are independent, i.e. $\pi(\vth) = \pi(\vth_{\rm L})\pi(\vth_{\rm NL})$
    \item The linear prior is either Gaussian or improper uniform.
\end{enumerate}
We note that in this work we consider both the linear and non-linear priors to be improper uniform for simplicity, and thus they drop out of our derivations. 
We furthermore note that our formulation in which $\vect{r}$ is the residual of raw data to a model, while being sufficiently general, is not the only way -- nor always the most practical -- to formulate the residuals. 
Indeed, the `data' $\vect{d}$ may be taken to be some function of the raw data, $\vect{d} = f(\vect{d}_{\rm raw}, \vth_{\rm NL})$, so long as the resulting data has a Gaussian distribution. In practice, the only realistic non-trivial function that preserves Gaussianity is a scaling, i.e. $\vect{d} = f(\vth_{\rm NL}) \vect{d}_{\rm raw}$. In this case, the parameters used in $f$ must be considered `non-linear' as they affect the covariance of $\vect{d}$. Given the constraints, this is mathematically equivalent to keeping $\vect{d} = \vect{d}_{\rm raw}$ and dividing the model $\vect{m}$ by $f(\vth_{\rm NL})$, in which case the covariance is constant (but the parameters must still be non-linear as they are divisors in the model). While mathematically equivalent, the two are not algorithmically equivalent, and which is computed more efficiently depends on the way a particular code is written.

We note that these conditions hold for our likelihood, \autoref{eq:field_likelihood_conceptual} (cf. Eqs. \ref{eq:dcal_gaussian} and \ref{eq:rsml}). 

We now integrate the posterior over the linear parameters:
\begin{align}
    p_{\rm NL}(\vth_{\rm NL}|\vect{d}) &= \int p(\vth_{\rm NL}, \vth_{\rm L}) d\vth_{\rm L} \\
    &\propto \pi_{\rm NL}(\vth_{\rm NL}) \int \pi_{\rm L}(\vth_{\rm L}) \mathcal{L}(\vect{d} | \vth_{\rm NL}, \vth_{\rm L}) d\vth_{\rm L} \\
    &= \int \mathcal{L}(\vect{d} | \vth_{\rm NL}, \vth_{\rm L}) d\vth_{\rm L} \\
    &\equiv \mathcal{L}_{\rm eff}(\vect{d} | \vth_{\rm NL}).
\end{align}
Here the second last equality makes the assumption that the priors on both linear and non-linear parameters are improper uniform distributions and the last equality defines an ``effective'' likelihood.

Let $m'$ be the model conditioned on the non-linear parameters. We can then write $m' = \matr{A} \vth_{\rm L}$, i.e. the remaining model is a linear model with design matrix $\matr{A}$ (note that $\matr{A}$ may be dependent on the non-linear parameters). Further, let the data covariance (after any transformation by non-linear parameters) be $\matr{\Sigma}$ and the maximum-likelihood estimate of the linear parameters (conditional on the chosen non-linear parameters) be $\hat{\vth}_{\rm L}$, that is,
\begin{equation}
    \hat{\vth}_{\rm L} = (\matr{A}^T \matr{\Sigma}^{-1} \matr{A})^{-1} \matr{A}^T \matr{\Sigma}^{-1} \vect{d} = \matr{\Sigma}_{\rm L} \matr{A}^T \matr{\Sigma}^{-1} \vect{d},
\end{equation}
with $\matr{\Sigma}_{\rm L}$ the covariance matrix of the linear parameters.
Now, express the residuals as $\vect{r} = \vect{d} - \matr{A} (\hat{\vth}_{\rm L} + \vect{\delta}_{\rm L}) \equiv \vect{\hat{r}} - \matr{A}\vect{\delta}_{\rm L} $, i.e. the sum of the maximum-likelihood residuals and a small model component.

Following the derivation in \citet{Monsalve2018}, we express the exponent of the conditional likelihood (not the effective likelihood) as
\begin{equation}
    -\frac{1}{2} \vect{\hat{r}}^T \matr{\Sigma}^{-1} \vect{\hat{r}} - \frac{1}{2} \vect{\delta}_{\rm L}^T \matr{A}^T \matr{\Sigma}^{-1} \matr{A} \vect{\delta}_{\rm L} + \vect{\hat{r}}^T \matr{\Sigma}^{-1} \matr{A} \vect{\delta}_{\rm L}.
\end{equation}
We may then integrate over the linear parameters:
\begin{equation}
    \mathcal{L}_{\rm eff}(\vect{d}|\vth_{\rm NL}) \propto \frac{\exp\left\{-\frac{1}{2} \vect{\hat{r}}^T \matr{\Sigma}^{-1} \vect{\hat{r}}\right\}} {\sqrt{|\matr{\Sigma}|}} \int \exp\left\{ -\frac{1}{2} \chi^2_\dagger \right\}\, d\vec{\delta}_{\rm L},
\end{equation}
where
\begin{equation}
    -\frac{1}{2} \chi^2_\dagger = -\frac{1}{2} \vth_{\rm L}^T \matr{A}^T \matr{\Sigma}^{-1} \matr{A} \vth_{\rm L} + \vect{d}^T \matr{\Sigma}^{-1} \matr{A} \vth_{\rm L}.
\end{equation}
This may be solved using the identity Eq. 13 in \citet{Monsalve2018} to give
\begin{equation}
    \int \exp\left\{ -\frac{1}{2} \chi^2_\dagger \right\}\, d\vect{\delta}_{\rm L} =  \sqrt{(2\pi)^{N_{\rm L}} |\matr{\Sigma}_{\rm L}|} \exp\left\{ \frac{1}{2} \vect{b}^T \matr{\Sigma}_{\rm L} \vect{b} \right\},
\end{equation}
with $\vect{b} = \matr{A}^T \matr{\Sigma}^{-1} \vect{\hat{r}}$.
Substituting this result back into the effective likelihood we have
\begin{align}
    \mathcal{L}_{\rm eff}(\vect{d}|\vth_{\rm NL}) &\propto \frac{\exp\left\{-\frac{1}{2} \vect{\hat{r}}^T \matr{\Sigma}^{-1} \vect{\hat{r}} + \frac{1}{2} \vect{b}^T \matr{\Sigma}_{\rm L} \vect{b} \right\}} {\sqrt{|\matr{\Sigma}| |\matr{\Sigma}^{-1}_{\rm L}|}} \\
    &= \frac{\exp\left\{-\frac{1}{2} \vect{\hat{r}}^T \matr{\Sigma}^{-1} \vect{\hat{r}} + \frac{1}{2} \vect{\hat{r}}^T \matr{\Sigma}^{-1} \matr{A} \matr{\Sigma}_{\rm L} \matr{A}^T \matr{\Sigma}^{-1} \vect{\hat{r}} \right\}} {\sqrt{|\matr{\Sigma}| |\matr{\Sigma}^{-1}_{\rm L}|}} \\
    &= \frac{\exp\left\{-\frac{1}{2} \vect{\hat{r}}^T \matr{\Sigma}^{-1} \left[\vect{\hat{r}} - \matr{A} \matr{\Sigma}_{\rm L} \matr{A}^T \matr{\Sigma}^{-1} \vect{\hat{r}} \right] \right\}} {\sqrt{|\matr{\Sigma}| |\matr{\Sigma}^{-1}_{\rm L}|}}.
\end{align}
Now, departing from the derivation of \citet{Monsalve2018}, we note that the last term in the exponential contains the standard ``hat matrix'',
\begin{equation}
    \matr{H} = \matr{A} \matr{\Sigma}_{\rm L} \matr{A}^T \matr{\Sigma}^{-1},
\end{equation}
which ``puts a hat'' on the data model, i.e. $\matr{H} \vect{d} = \matr{A} \hat{\vth}_{\rm L}$ and is idempotent. We thus have 
\begin{align}
    \mathcal{L}_{\rm eff}(\vect{d}|\vth_{\rm NL}) &\propto \frac{\exp\left\{-\frac{1}{2} \vect{\hat{r}}^T \matr{\Sigma}^{-1} \left[\vect{\hat{r}} - \matr{H} (\vect{d} - \matr{A} \hat{\vth}_{\rm L}) \right] \right\}} {\sqrt{|\matr{\Sigma}| |\matr{\Sigma}^{-1}_{\rm L}|}} \\
    &= \frac{\exp\left\{-\frac{1}{2} \vect{\hat{r}}^T \matr{\Sigma}^{-1} \left[\vect{\hat{r}} -  (\matr{H} \vect{d} - \matr{H} \matr{H} \vect{d}) \right] \right\}} {\sqrt{|\matr{\Sigma}| |\matr{\Sigma}^{-1}_{\rm L}|}} \\
    &= \frac{\exp\left\{-\frac{1}{2} \vect{\hat{r}}^T \matr{\Sigma}^{-1} \vect{\hat{r}} \right\}} {\sqrt{|\matr{\Sigma}| |\matr{\Sigma}^{-1}_{\rm L}|}},
    \label{eq:marginalized-likelihood}
\end{align}
where the last equality follows due to idempotency of $\matr{H}$.

This final equation, \cref{eq:marginalized-likelihood}, is the same result as given in \citet{Tauscher2021}.

\section{Linear Representation of Calibration Likelihood}
\label{app:linear-cal-model}
In \S\ref{sec:cal:naive} we presented the likelihood of the calibration data in a clear conceptually-oriented notation.
In practice, to use the analytical marginalization over the linear parameters, as outlined in \S\ref{sec:amlp} and App. \ref{app:marginalized-likelihood}, it is useful to represent the data model as a combination of linear and non-linear parameters, in which the the linear parameters enter exclusively through a single term, $\matr{A} \vth_{\rm L}$.
Here, we derive this representation.

Let the linear temperature terms, $T_{\mathcal{T}_{\rm lin}}$, be formed into a vector
\begin{equation}
    \vth_{\rm NW+L} = \left[ \vth_{\rm unc}^T, \vth_{\rm cos}^T,\vth_{\rm sin}^T,\vth_{\rm L}^T \right]^T.
\end{equation}
Also, for a particular input source, ${\rm src} \in \mathcal{S}_{\rm cal}$, and term, $p \in \mathcal{T}_{\rm lin}$, define an $N_\nu \times N^p_{\rm terms}$ sub-design matrix 
\begin{equation}
    \matr{V}^{{\rm src},p}_{ij} = \vect{\kappa}^{\rm src}_p (\vect{\nu}_i) \matr{\Psi}^p_{ij}.
    \label{eq:matrixv}
\end{equation}
Then define a design matrix:
\begin{equation}
    \matr{K}_{\rm cal} = \begin{pmatrix}
        \matr{V}_{\rm amb, unc} & \matr{V}_{\rm amb, cos} & \matr{V}_{\rm amb, sin} & \matr{V}_{\rm amb, L} \\
        \matr{V}_{\rm hot, unc} & \matr{V}_{\rm hot, cos} & \matr{V}_{\rm hot, sin} & \matr{V}_{\rm hot, L} \\
        \matr{V}_{\rm short, unc} & \matr{V}_{\rm short, cos} & \matr{V}_{\rm short, sin} & \matr{V}_{\rm short, L} \\
        \matr{V}_{\rm open, unc} & \matr{V}_{\rm open, cos} & \matr{V}_{\rm open, sin} & \matr{V}_{\rm open, L} 
    \end{pmatrix}
    \label{eq:matrix_K}
\end{equation}

Furthermore, let the LHS of \autoref{eq:per-source-linear} for a particular input source be written as a $N_\nu$-vector $\vect{d}_{\rm src}$\footnote{We choose the notation $\vect{d}$ here as this quantity represents our `data', though in truth it is a linear transformation of the data, in which the transformation itself is being modelled.}:
\begin{equation}
    \vect{d}_{\rm src} = \vect{q}_{\rm src} \circ \matr{\Psi} \vth_{\rm NS} - \vect{\rho}_{\rm src} \circ \vect{T}_{\rm src},
    \label{eq:dsrc}
\end{equation}
and form the $N_\nu |\mathcal{T}_{\rm cal}|$ vector concatenated over sources:
\begin{equation}
    \vect{d}_{\rm cal} = \left[ \vect{d}^T_{\rm amb}, \vect{d}^T_{\rm hot}, \vect{d}^T_{\rm open}, \vect{d}^T_{\rm short} \right]^T.
    \label{eq:dcal}
\end{equation}

Under the approximations of Gaussianity and independence of frequency channels, we can write the model residual vector $\vect{r}_{\rm cal}$ in the following form (equivalent to the concatenation of $\vect{r}_{\rm src}$ vectors, cf. Eq. \ref{eq:rsrc}):
\begin{equation}
    \vect{r}_{\rm cal} = \left(\vect{d}_{\rm cal} - \matr{K}_{\rm cal} \vth_{\rm NW+L}\right)  \sim  \mathcal{N}(0, \matr{\Sigma}_{\rm cal}),
    \label{eq:dcal_gaussian}
\end{equation}
where $\matr{\Sigma}_{\rm cal}$ is the diagonal scaled covariance matrix given by 
\begin{equation}
    \matr{\Sigma}_{\rm cal} = \begin{pmatrix}
        \matr{\Sigma}_{\rm amb} & 0 & 0 & 0 \\
         0& \matr{\Sigma}_{\rm hot} & 0 & 0\\
         0& 0& \matr{\Sigma}_{\rm short} &0  \\
         0&0 & 0 & \matr{\Sigma}_{\rm open} 
    \end{pmatrix}.
    \label{eq:sigmacal}
\end{equation}

Eq. \ref{eq:dcal_gaussian} represents a Gaussian likelihood for the (transformed) data $\vect{d}_{\rm cal}$ given a purely linear model $\matr{K}_{\rm cal}\vth_{\rm NW+L}$ in which $\matr{K}_{\rm cal}$ in general depends on non-linear parameters (but we do not consider any such parameters in this paper). This is precisely the form we need to apply the effective likelihood, Eq. \ref{eq:marginalized-likelihood}, with $\matr{A} = \matr{K}_{\rm cal}$.
Note that we have adopted the scaled Gaussian likelihood (Eq. \ref{eq:general-gaussian-likelihood}) in which both $\vect{d}_{\rm cal}$ and $\matr{\Sigma}_{\rm cal}$ are scaled by the non-linear parameters $\vth_{\rm NS}$.

\section{Linear Representation of Joint Likelihood}
\label{app:linear-field-model}
Here, we apply the same process to the joint likelihood, Eq. \ref{eq:field_likelihood_conceptual}, as App. \ref{app:linear-cal-model} applied to the calibration likelihood, Eq. \ref{eq:cal_likelihood_conceptual}. Our aim is to represent the likelihood in the same form as \ref{eq:dcal_gaussian}.

We treat the antenna simply as another (i.e. fifth) source. 
Thus we have 
\begin{equation}
    \vect{d}_{\rm ant} = \bar{\vect{q}}_{\rm ant} \circ \matr{\Psi} \vth_{\rm NS} - \vect{\rho}_{\rm ant} \circ \vect{T}_{\rm {21, meas}},
\end{equation}
where
\begin{equation}
    \vect{T}_{\rm 21, meas} = \vect{L} \circ \bar{\vect{b}}_{\rm corr} \circ \vect{T}_{21}(\vth_{21}) +
    \overline{\vect{T}}_{\rm loss}
\end{equation}
and
\begin{equation}
    \vect{d}_{\rm sml} = \left[ \vect{d}^T_{\rm amb}, \vect{d}^T_{\rm hot}, \vect{d}^T_{\rm open}, \vect{d}^T_{\rm short}, \vect{d}^T_{\rm ant}\right]^T.
    \label{eq:dsml}
\end{equation}
Note that this has removed the contribution of the foregrounds to the expected antenna temperature, as these are linear and will be added to the linear term, rather than $\vect{d}$.
To do this, we modify the $\matr{K}$ matrix to be
\begin{equation}
    \matr{K}_{\rm sml} = \begin{pmatrix}
        \matr{V}_{\rm amb, unc} & \matr{V}_{\rm amb, cos} & \matr{V}_{\rm amb, sin} & \matr{V}_{\rm amb, L} & 0\\
        \matr{V}_{\rm hot, unc} & \matr{V}_{\rm hot, cos} & \matr{V}_{\rm hot, sin} & \matr{V}_{\rm hot, L} & 0  \\
        \matr{V}_{\rm short, unc} & \matr{V}_{\rm short, cos} & \matr{V}_{\rm short, sin} & \matr{V}_{\rm short, L} & 0\\
        \matr{V}_{\rm open, unc} & \matr{V}_{\rm open, cos} & \matr{V}_{\rm open, sin} & \matr{V}_{\rm open, L} & 0 \\
        \matr{V}_{\rm ant, unc} & \matr{V}_{\rm ant, cos} & \matr{V}_{\rm ant, sin} & \matr{V}_{\rm ant, L} & 
        \matr{V}_{\rm ant, FG}
    \end{pmatrix},
    \label{eq:matrix_K_full}
\end{equation}
where
\begin{equation}
    \matr{V}^{\rm ant, FG}_{ij} = \rho_{i, {\rm ant}} \vect{L}_{i} \vect{\bar{b}}_{\rm corr} \matr{\Phi}_{ij},    
\end{equation}
and write $\vth_{\rm lin}$ as 
\begin{equation}
    \vth_{\rm lin} = \left[ \vth_{\rm unc}^T, \vth_{\rm cos}^T,\vth_{\rm sin}^T,\vth_{\rm L}^T , \vth_{\rm FG}^T\right]^T.
\end{equation}

Under the same assumptions and arguments employed in App. \ref{app:linear-cal-model}, this yields
\begin{equation}
    \vect{r}_{\rm sml} = \vect{d}_{\rm full}(\vth_{21}, \vth_{\rm NS}) -  \matr{K}_{\rm sml} \vth_{\rm lin} \sim \mathcal{N}(0, \matr{\Sigma}_{\rm sml}),
    \label{eq:rsml}
\end{equation}
with $\matr{\Sigma}_{\rm sml}$ is simply
\begin{equation}
    \matr{\Sigma}_{\rm cal} = \begin{pmatrix}
        \matr{\Sigma}_{\rm amb} & 0 & 0 & 0 & 0 \\
         0& \matr{\Sigma}_{\rm hot} & 0 & 0& 0  \\
         0& 0& \matr{\Sigma}_{\rm short} &0 & 0 \\
         0&0 & 0 & \matr{\Sigma}_{\rm open} & 0 \\
         0&0 & 0 & 0 & \matr{\Sigma}_{\rm ant}  \\
    \end{pmatrix}.
    \label{eq:sigmasml}
\end{equation}

\section{Obtaining `Recalibrated' Sky Temperature}
\label{sec:recal}
Given calibrated sky temperature data, $\widehat{\overline{T}}_{\rm sky,bc}$ that was calibrated with the multiplicative and additive temperatures $\widehat{T}_0^{\rm ant}$ and $\widehat{T}^{\rm ant}_1$ respectively, the data may be re-calibrated using new estimates of the calibration temperatures,  $\widehat{T}^{\rm ant'}_0$ and $\widehat{T}^{\rm ant'}_1$ as follows:
\begin{align}
    \widehat{\overline{T}}'_{\rm sky,bc} &= \frac{1}{L\bar{b}_{\rm corr}}\left[x_0 \left(L \bar{b}_{\rm corr} \widehat{\overline{T}}_{\rm sky,bc} + \overline{T}_{\rm loss} - \widehat{T}_1^{\rm ant}\right) + \widehat{T}_{1}^{\rm ant'} - \overline{T}_{\rm loss} \right] \nonumber \\
    &= x_0 \widehat{\overline{T}}_{\rm sky,bc} + \frac{1}{L\bar{b}_{\rm corr}}\left[\overline{T}_{\rm loss}\left(x_0 - 1\right) - x_0 \widehat{T}_1^{\rm ant} + \widehat{T}_1^{\rm ant'} \right]
    \label{eq:recalibrate}
\end{align}
where $x_0 = \widehat{T}_0^{\rm ant'} / \widehat{T}_0^{\rm ant}$ is the ratio of the new to old scaling temperatures. 
Here, $L$, $\bar{b}_{\rm corr}$ and $\overline{T}_{\rm loss}$ are defined in \S\ref{sec:field-data-model}.
This re-calibrated temperature is used in Fig. \ref{fig:differences_with_alan}.

\bsp	
\label{lastpage}
\end{document}